\documentclass[aps,pra,twocolumn,
showpacs,nofootinbib]{revtex4-1}
\usepackage{graphicx}
\usepackage{color}
\usepackage{blindtext}
\usepackage{amsmath}
\usepackage{amsfonts}
\usepackage{braket}

\DeclareMathOperator{\sech}{sech}
\DeclareMathOperator{\csch}{csch}
\newcommand{\dash}{-}
\newcommand{\Lagr}{\mathcal{L}}

\begin{document}

\title{Non-integrable dynamics of matter-wave solitons in a
density-dependent gauge theory}
\author{R. J. Dingwall$^1$}
\author{M. J. Edmonds$^{2,3}$}
\author{J. L. Helm$^{4,5}$}
\author{B. A. Malomed$^6$}
\author{P. \"{O}hberg$^1$}
\affiliation{$^1$SUPA, Institute of Photonics and Quantum Sciences,
Heriot-Watt University, Edinburgh, EH14 4AS, United Kingdom}
\affiliation{$^2$Joint Quantum Centre Durham-Newcastle,
School of Mathematics and Statistics, Newcastle University,
Newcastle upon Tyne, NE1 7RU, United Kingdom}
\affiliation{$^3$Quantum Systems
Unit, Okinawa Institute of Science and Technology Graduate University,
Okinawa 904-0495, Japan}
\affiliation{$^4$Joint Quantum Centre Durham-Newcastle, Department of Physics,
Durham University, Durham, DH1 3LE, United Kingdom}
\affiliation{$^5$Department
of Physics, University of Otago, Dunedin, New Zealand}
\affiliation{$^6$Department of Physical Electronics, School of Electrical Engineering,
Faculty of Engineering, Tel Aviv University, Tel Aviv 69978, Israel}

\begin{abstract}
We study interactions between bright matter-wave solitons which acquire
chiral transport dynamics due to an optically-induced density-dependent
gauge potential. Through numerical simulations, we find that the collision
dynamics feature several non-integrable phenomena, from inelastic collisions
including population transfer and radiation losses to short-lived bound
states and soliton fission. An effective quasi-particle model for the
interaction between the solitons is derived by means of a variational
approximation, which demonstrates that the inelastic nature of the collision
arises from a coupling of the gauge field to velocities of the solitons. In
addition, we derive a set of interaction potentials which show that the
influence of the gauge field appears as a short-range potential, that can
give rise to both attractive and repulsive interactions.
\end{abstract}

\maketitle



\section{\label{sec:level1}Introduction}

One of the defining properties of solitons in systems such as the nonlinear
Schr\"{o}dinger (NLS) and Korteweg-de Vries equations is that they pass
through and emerge from collisions with other solitons unperturbed, with the
exception of a phase shift arising from the nonlinear interaction. \cite%
{tome,solbook}. All of the dynamical quantities of solitons, such as their
velocities and masses, are conserved during the collision. In such systems,
the elastic nature of these collisions is a consequence of the integrability
of the model, which heavily restricts the allowed dynamics due to the
existence of an infinite set of conservation laws.

In non-integrable systems, stable solitons may exist too, but their
collisions are, generally, inelastic and can lead to trajectories
which are chaotic \cite{tome,solbook2}. In this case, the defining feature
is the existence of short-lived bound states in which the number of
collision events depends fractally on the initial conditions. Generally,
this mechanism arises from the excitation of an internal mode of the
solitons in non-integrable models, either with \cite%
{chaos5,chaos6,chaos7,chaos10,chaos11} or without radiation losses \cite%
{chaos8,chaos9}, or through the presence of a weak perturbation \cite%
{chaos1,chaos2,chaos4}. Solitons can merge or fracture into new
products through fission and fusion processes \cite{fusfis1,fusfis2,fusfis3}%
, which has also been studied in the context of three-soliton and
soliton-breather collisions \cite{boris3}. These effects highlight a strong
contrast to soliton dynamics in integrable systems, which are not only
interesting from a fundamental point of view, but offer insight into the
description of realistic systems where the influence of perturbations can be
consequential.

Ultracold atomic gases represent an attractive platform to study nonlinear
physics due to their unprecedented experimental controllability. The ability
to precisely engineer both the dimensionality and interactions in these
systems has lead to the realization of isolated bright matter-wave solitons
\cite{sol1,bsw2}, as well as solitons trains \cite{sol2}. The second
generation of experiments addressed controlled collisions with potential
barriers \cite{marchant_2013,marchant_2016}, as well as understanding both
the role and origin of the relative phase for the stability of bright
soliton states \cite{trapcoldcol1,nguyen_2017,everitt_2017}.

The underlying integrability of the focusing nonlinear Schr\"{o}dinger
equation leads to a hierarchy of analytical higher-order soliton solutions
provided by the inverse scattering transform. This directly led to the
development of a classical particle model \cite{trapcoldcol3}, describing
the dynamics of the bright solitons, from which regions of chaotic behaviour
have been eventually predicted for trapped bright solitons in non-integrable
settings \cite{trapcoldcol4,martin_2008}.

Due to their inherent coherence, bright solitons represent a useful tool for
investigating interferometry in the quantum realm. Recent experimental
progress in this direction has seen the first realization of a $^{85}$Rb
matter-wave Interferometer, as well as theoretical proposals for precision
measurement using the Sagnac effect \cite{helm_2015} and the creation of
Bell states using quantum bright solitons \cite{gertjerenken_2013}, as well
taking advantage of the interaction of solitons with nonlinear splitters
\cite{HS}.

The ability to simulate artificial gauge theories with ultracold gases
offers a new opportunity to understand the interplay of effective magnetism
in these systems with nonlinear effects \cite{dalibard_2011,Lightgauge}.
This has led to the realization of vortex states \cite{lin_2009} as well as
spin-orbit coupling \cite{lin_2011}. Interest was recently focused on
schemes for generating gauge potentials with an effective back-action
between the matter and the gauge potential \cite{intgauge,intgauge3}, which
leads to a number of novel phenomena, including the violation of the Kohn's
theorem \cite{matt1,intgauge2s}, and unconventional vortex dynamics \cite%
{giulio1,giulio2}. Very recently, the first experimental realization of a
dynamical gauge theory in a trapped ion system was shown \cite{martinez_2016}%
.

In this paper, we study the nonlinear dynamics of two interacting
one-dimensional chiral matter-wave solitons. We begin by reviewing how these
solitons can be engineered in ultracold gases using optical techniques which
induce an effective density-dependent gauge potential in the atomic cloud.
The resulting equation of motion for the gas, which takes the form of a
chiral nonlinear Schr\"{o}dinger equation, is then solved numerically in
Sec. \ref{Numerics}. Following this in Sec. \ref{Variational}, we develop a
variational approach to further understand the soliton dynamics in this
system, both in linear and asymptotic limits, before concluding in Sec. \ref%
{Conclusion}.

\section{\label{sec:level1}The theoretical model}

We consider a Bose-Einstein condensate of $N$ two-level interacting atoms,
in which two internal states of the atoms (labelled $\ket{1}$ and $\ket{2}$)
are resonantly coupled by an external laser field. The Hamiltonian
describing the interacting trapped gas as well as the light-matter
interaction can be written as
\begin{equation}
\hat{H}=\bigg(\frac{\hat{\mathbf{p}}^{2}}{2m}+\frac{1}{2}m\omega _{\perp
}^{2}\mathbf{r}_{\mathrm{\perp }}^{2}\bigg)\otimes \mathbb{I}+\hat{H}_{%
\mathrm{lm}}+\hat{\mathcal{V}}_{\mathrm{int}},  \label{Hammy}
\end{equation}%
where
\begin{equation}
\hat{H}_{\mathrm{lm}}=\frac{\hbar \Omega }{2}\bigg(%
\begin{array}{cc}
0 & e^{-i\phi _{\ell }} \\
e^{i\phi _{\ell }} & 0%
\end{array}%
\bigg)  \label{eqn:hlm}
\end{equation}%
describes the optical coupling of the two internal states of the atoms, with
strength $\Omega $ and laser phase $\phi _{\ell }$. In order to obtain an
equation of motion for the many-particle system, the state of the system is
defined as a Hartree product $\ket{\Psi}=\otimes _{l=1}^{N}%
\ket{\chi_{l}^{(0)}}$, where $\ket{\chi_{l}^{(0)}}$ defines one of the
single-particle eigenstates of Eq.~\eqref{eqn:hlm}. In this work we assume
that the gas is harmonically trapped in the $\left( x,y\right) $ plane
(described by the vectorial $\mathbf{r}_{\perp }$ coordinates in Eq.~%
\eqref{eqn:hlm}), but free along the axial $z$ direction. The mean-field
interactions appearing in Eq.~\eqref{Hammy} are defined by $\hat{\mathcal{V}}%
_{\mathrm{int}}=\text{diag}[g_{11}|\Phi _{1}|^{2}+g_{12}|\Phi
_{2}|^{2},g_{22}|\Phi _{2}|^{2}+g_{12}|\Phi _{1}|^{2}]$, where $|\Phi
_{i}|^{2}$ denotes the population of the state $i$.

Provided that the gas is sufficiently dilute, we can diagonalize the
Hamiltonian by treating the mean-field interactions $g_{ii^\prime}\vert%
\Phi_i\vert^2$ as a small perturbation to the laser coupling $\hbar\Omega$.
The eigenvectors of Eq. (\ref{Hammy}) can then be written in the dressed
state basis $\{+,-\}$ as
\begin{equation}  \label{dressedstates}
\vert\chi_\pm\rangle=\vert\chi^{(0)}_\pm\rangle+\frac{g_{11}-g_{22}}{%
8\hbar\Omega}\vert\Phi_\pm\vert^2\vert\chi^{(0)}_\mp\rangle,
\end{equation}
where $\vert\chi^{(0)}_\pm\rangle=\left(\vert 1\rangle\pm
e^{i\phi_\ell}\vert 2\rangle\right)/\sqrt{2}$ denotes the unperturbed
dressed states. The associated eigenvalues are given by $g\vert\Phi_\pm%
\vert^2\pm\hbar\Omega/2$, with the dressed scattering parameter $%
g=(g_{11}+g_{22}+2g_{12})/4$.

From the interacting dressed states, we can write the state vector of the
system as $\ket{\xi}=\sum_{i=+,-}\Phi _{i}\left( \mathbf{r},t\right) |\chi
_{i}\rangle $. Then, the effective Hamiltonian is written as
\begin{equation}
\hat{H}_{\pm }=\frac{1}{2m}(\mathbf{p}-\mathbf{A}_{\pm })^{2}+\frac{1}{2}%
m\omega _{\perp }^{2}\mathbf{r}_{\perp }^{2}+\frac{g}{2}|\Phi _{\pm }|^{2}.
\label{eqn:hf}
\end{equation}%
Equation \eqref{eqn:hf} introduces the geometric phase $\mathbf{A}_{\pm
}=i\hbar \langle \chi _{\pm }|\nabla \chi _{\pm }\rangle $. Accompanying
this is a scalar geometric phase, whose leading-order effect is inducing an
energy offset, which may be dropped. Then, using the definition given by Eq.~%
\eqref{dressedstates}, to lowest-order the density-dependent geometric phase
appears as $\mathbf{A}_{\pm }=\mathbf{A}^{(0)}+\mathbf{a}_{1}|\Phi _{\pm }(%
\mathbf{r})|^{2}$, with the single-particle vector potential, $\mathbf{A}%
^{(0)}=-\frac{\hbar }{2}\nabla \phi _{l}(\mathbf{r})$, while $\mathbf{a}%
_{1}=\nabla \phi _{l}(\mathbf{r})(g_{11}-g_{22})/8\Omega $ defines the
strength of the density-dependent gauge potential. The equation of motion
governing the evolution of the wave-function amplitude $\Phi _{+}\left(
\mathbf{r},t\right) $ is found from minimization of the system's energy
functional, $\mathcal{E}=\langle \Psi |(i\hbar \partial _{t}-\hat{H}_{\pm
})|\Psi \rangle $. After dropping $\pm $ subscripts, the resulting
mean-field density-dependent Gross-Pitaevskii equation is obtained as
\begin{equation}
i\hbar \frac{\partial \Phi }{\partial t}=\left[ \frac{1}{2m}\left( \mathbf{p}%
{-}\mathbf{A}\right) ^{2}{+}\mathbf{a_{1}}\cdot \mathbf{j}{+}\frac{1}{2}%
m\omega _{\perp }^{2}\mathbf{r}_{\perp }^{2}{+}g|\Phi |^{2}\right] \Phi ,
\label{3DEOM}
\end{equation}%
%
%
where
\begin{equation}
\mathbf{j}=\frac{1}{2m}\bigg[\Phi \left( \mathbf{p}+\mathbf{A}\right) \Phi
^{\ast }-\Phi ^{\ast }\left( \mathbf{p}-\mathbf{A}\right) \Phi \bigg].
\label{3Dcurrent}
\end{equation}%
defines the current nonlinearity appearing in Eq.~\eqref{3DEOM}. The
current-coupled nonlinear Schr\"{o}dinger equation, captured by Eq.~%
\eqref{3DEOM}, describes a novel nonlinear gauge theory where there is an
effective back-action between the matter-field and the gauge potential \cite%
{intgauge}. This feedback ingredient of the system is somewhat similar to
the local field effect, which affects a \textquotedblleft soft" optical
lattice trapping the condensate, that gives rise to various consequences,
such as formation of bright solitons in the absence of contact interactions
between atoms \cite{GDong}.

\subsection{\label{sec:level2}One-dimensional reduction}

We are interested in studying solitary-wave solutions in the frameworks of
the dimensionally reduced form of Eqs.~\eqref{3DEOM} and \eqref{3Dcurrent}.
To do this, we assume the system is in the ground state of the transverse
trap, such that one can factorize the wave function as $\Phi (\mathbf{r}%
,t)=\Psi _{\perp }(\mathbf{r}_{\perp })\Psi (x,t)$, where $\Psi _{\perp }(%
\mathbf{r}_{\perp })=(\sqrt{\pi }l_{\perp })^{-1}\exp (-\mathbf{r}_{\perp
}^{2}/2l_{\perp }^{2})$ is the transverse ground-state wave function, and $%
l_{\perp }=\sqrt{\hbar /m\omega _{\perp }}$ is the transverse harmonic
length scale. Equation (\ref{3DEOM}) is then reduced to an effective
one-dimensional form,
\begin{equation}
i\hbar \frac{\partial \Psi }{\partial t}=\left[ \frac{1}{2m}\left( \hat{p}%
-a_{1}|\Psi |^{2}\right) ^{2}+a_{1}j(x)+g_{\mathrm{1D}}|\Psi |^{2}\right]
\Psi ,  \label{mincoupleEOM}
\end{equation}%
where both the scattering parameters $a_{1}=k(g_{11}-g_{22})/(16\pi l_{\perp
}^{2}\Omega )$ and $g_{\mathrm{1D}}=g/(2\pi l_{\perp }^{2})$ have been
scaled by the transverse area of the cloud, and the corresponding
one-dimensional current nonlinearity is defined as
\begin{equation}
j(x)=\frac{1}{2m}\bigg[\Psi \left( \hat{p}{+}a_{1}|\Psi |^{2}\right) \Psi
^{\ast }{-}\Psi ^{\ast }\left( \hat{p}{-}a_{1}|\Psi |^{2}\right) \Psi \bigg].
\label{eqn:j1d}
\end{equation}%
In writing Eq.~\eqref{mincoupleEOM}, we have defined the laser phase as $%
\phi _{\ell }=kx$ and subsequently eliminated the zeroth-order vector
potential through a momentum boost. We can further simplify Eq. (\ref%
{mincoupleEOM}) by introducing the nonlinear phase transformation
\begin{equation}
\Psi (x,t)=\psi (x,t)\text{ exp}\left( \frac{ia_{1}}{\hbar }\int_{-\infty
}^{x}dx^{\prime }|\psi (x^{\prime },t)|^{2}\right) ,  \label{nonlineartrans}
\end{equation}%
which acts to decouple the vector potential from the canonical momentum
appearing in the one-dimensional Gross-Pitaevskii equation. Substituting Eq.~%
\eqref{nonlineartrans} into Eq.~\eqref{mincoupleEOM} and \eqref{eqn:j1d}
leads to the simplified equation,
\begin{equation}
i\hbar \frac{\partial \psi }{\partial t}=\bigg[-\frac{\hbar ^{2}}{2m}%
\partial _{x}^{2}-2a_{1}j^{\prime }(x)+g_{\mathrm{1D}}|\psi |^{2}\bigg]\psi ,
\label{gaugeEOM}
\end{equation}%
where $j^{\prime }(x)=(\hbar /m)\text{Im}(\psi ^{\ast }\partial _{x}\psi )$
is the gauge-transformed current operator.
Equation~\eqref{gaugeEOM} belongs to the class of derivative or `chiral' NLS
equations \cite{derivNLS,chiral1}, which was originally studied, in
particular, in the context of one-dimensional anyons \cite{anyon2}. The
model features several key differences from the standard NLS equation in
that it is generally non-integrable, does not obey the Galilean invariance,
and possesses chiral soliton solutions \cite{anyon,anyon2,chiral1}. These
properties are expected to contribute to unconventional soliton dynamics in
the one-dimensional case.

An experimental realization of the interacting gauge theory relies,
basically, on two conditions, $viz$., an atomic species possessing
long-lived excited states that the adiabatic motion of the atoms requires,
as well as the spontaneous-emission rate that is negligible on the time
scale of cold-atom experiments. A promising candidate that fulfils these
conditions are alkali-earth atoms, which have been recently used to create a
spin-orbit-coupled Fermi gas of $^{173}$Yb atoms \cite{song_2016}, using a
methodology similar to that outlined here. Very recently, an interaction-induced synthetic gauge potential was realised experimentally in a Bose-Einstein condensate loaded into a  modulated two-dimensional lattice \cite{Clark_2018}.

\subsection{\label{sec:level2}Chiral solitons}

Single-soliton solutions of Eq. (\ref{gaugeEOM}) can be derived by first
boosting into the moving frame via the transformation
\begin{equation}
\psi _{\mathrm{BS}}(x,t)=\varphi (x-vt)e^{i(mvx^{\prime }+mv^{2}t^{\prime
}/2-\mu t^{\prime })/\hbar },  \label{Galil1}
\end{equation}%
which consists of a Galilean transformation in which the stationary
coordinates $(x,t)$ and moving coordinates $(x^{\prime },t^{\prime })$ are
connected by the translations, $x^{\prime }\rightarrow x-vt$ and $t^{\prime
}\rightarrow t$, with frame velocity $v$. The resulting differential
equation for the real-valued wave function $\varphi (x^{\prime })$ becomes
\begin{equation}
\mu \varphi =-\frac{\hbar ^{2}}{2m}\frac{d^{2}}{d{x^{\prime }}^{2}}\varphi
+\left( g_{\mathrm{1D}}-2a_{1}v\right) \varphi ^{3},  \label{nls}
\end{equation}%
in which the current is contained as $j(x)=v\varphi ^{2}$. Integrating Eq. (%
\ref{nls}), and requiring that the wave function converges to $\varphi (\pm
\infty )=0$ for $g_{\mathrm{1D}}^{\prime }=g_{\mathrm{1D}}-2a_{1}v<0$, one
finds the single bright-soliton solution,
\begin{equation}
\psi _{\mathrm{BS}}={\frac{1}{\sqrt{2b}}}\sech\left( \left( x-vt\right)
/b\right) e^{i\left( mvx-mv^{2}t/2-\mu t\right) /\hbar },
\label{brightsoliton}
\end{equation}%
in which we have transformed back into the stationary frame. The chemical
potential appearing in Eq.~\eqref{nls} is $\mu =-mg_{\mathrm{1D}}^{2\prime
}N^{2}/8\hbar ^{2}$, with the amplitude factor $1/\sqrt{2b}$ provided normalization $N=1$. In
contrast to the Gross-Pitaevskii theory used to model non-chiral bright
solitons \cite{PandS}, the lack of the Galilean invariance of Eq. (\ref%
{gaugeEOM}) results in a change of the soliton width, $b=-2\hbar ^{2}/mg_{%
\mathrm{1D}}^{\prime }$, depending on the direction of the soliton's
motion. These solitons are therefore chiral, in the sense that the
mean-field interactions, and hence the soliton's size depend on the
direction in which it is travelling. An illustrative example of this can be
demonstrated by the reflection of a chiral soliton off a hard wall, which
causes the soliton to disperse \cite{intgauge}.

\subsection{\label{sec:level2}Conservation laws}

Although the equations of motion defined by Eqs. (\ref{mincoupleEOM}) and (%
\ref{gaugeEOM}) are non-integrable, a set of conservation laws in the
present system can be derived directly from the Noether's theorem \cite{tome}%
. With the respective Lagrangian density,
\begin{equation}
\mathcal{L}=\frac{i\hbar }{2}\left( \Psi \partial _{t}\Psi ^{\ast }-\Psi
^{\ast }\partial _{t}\Psi \right) +\frac{1}{2m}|\left( \hat{p}-a_{1}|\Psi
|^{2}\right) \Psi |^{2}+\frac{g_{\mathrm{1D}}}{2}|\Psi |^{4},
\label{lagdens}
\end{equation}%
it is straightforward to show that, at least, three conserved quantities
exist \cite{chiral6}, given by the integral expressions:
\begin{equation}
N=\int_{-\infty }^{\infty }dx|\Psi |^{2}=\int_{-\infty }^{\infty }dx|\psi
|^{2},  \label{density}
\end{equation}%
\begin{equation}
\begin{split}
P& =\frac{i\hbar }{2}\int_{-\infty }^{\infty }dx\left( \Psi ^{\ast }\partial
_{x}\Psi -\Psi \partial _{x}\Psi ^{\ast }\right) \\
& =-\int_{-\infty }^{\infty }dx\left( mj^{\prime }(x)+a_{1}|\psi
|^{4}\right) ,
\end{split}
\label{momentum}
\end{equation}%
and
\begin{equation}
\begin{split}
E& =-\int_{-\infty }^{\infty }dx\left( \frac{1}{2m}|(\hat{p}-a_{1}|\Psi
|^{2})\Psi |^{2}+\frac{g_{\mathrm{1D}}}{2}|\Psi |^{4}\right) \\
& =-\int_{-\infty }^{\infty }dx\left( \frac{\hbar ^{2}}{2m}|\partial
_{x}\psi |^{2}+\frac{g_{\mathrm{1D}}}{2}|\psi |^{4}\right) ,
\end{split}
\label{energy}
\end{equation}%
which correspond to the number of particles, momentum, and energy of the
system, respectively. The integrands of Eq. (\ref{energy}) introduce the
Hamiltonian densities for both the transformed and non-transformed
representations, which is expected due to the fact that the underlying
Lagrangian is Hermitian. A specific peculiarity of the chiral model arises
in Eq. (\ref{momentum}), which shows that, as a consequence of the breakdown
of the Galilean invariance, the canonical momentum is not conserved, but the
quantity $mj^{\prime }+a_{1}|\psi |^{4}$ is conserved.

\section{\label{Numerics}Numerical simulations}

The focussing NLS equation is an integrable model, where solitons collide
elastically, with the same shape and velocity before and after scattering.
The one-dimensional gauge theory defined by Eq.~\eqref{gaugeEOM} breaks the
integrability due to the presence of the current nonlinearity. In this
section we numerically solve Eq.~\eqref{gaugeEOM} for binary chiral-soliton
collisions using the known single-soliton solution, given by Eq.~%
\eqref{brightsoliton}, in order to understand how the broken integrability
manifests itself. The system is prepared initially in the state
\begin{equation}
\psi _{\mathrm{in}}(x,t{=}0)=\psi _{\mathrm{BS}}(x-\xi _{1})+e^{i\delta
}\psi _{\mathrm{BS}}(x-\xi _{2})  \label{eqn:bsz}
\end{equation}%
where $\delta \in \lbrack -\pi ,\pi ]$ is the relative phase difference
between the solitons, while $\xi _{1,2}$ are initial center-of-mass
coordinates of the solitons, and the normalization condition is $%
\int_{-\infty }^{+\infty }dx|\psi _{\mathrm{in}}(x)|^{2}=N$. Since a full
parameter scan of chiral-soliton collisions, featuring every degree of
freedom, presents a formidable problem, we restrict our analysis to two
parameter regimes, each set by a ratio of interaction strengths, which
illustrate the essential physics present in the model:

\begin{itemize}
\item In the first instance, we consider the case of strong chiral
interactions, $|g_{\mathrm{1D}}|\ll |a_{1}\left( v_{1}+v_{2}\right) |$,
where effects stemming from the current nonlinearity are made influential by
the solitons' high velocities.

\item For the second, we treat the case of weak chiral interactions, $|g_{%
\mathrm{1D}}|\gg |a_{1}\left( v_{1}+v_{2}\right) |$, where the current
nonlinearity is treated as a small perturbation added to the usual
mean-field dynamics.
\end{itemize}

To integrate Eq. (\ref{gaugeEOM}) numerically, we construct an explicit
central-difference scheme for the evolution of the wave function, and
compare the results to those produced by a split-step Fourier method, to
ensure consistency. The numerical domain is chosen to be at least two orders of magnitude larger than the widths of the solitons to avoid radiation back-reflecting into the solitons (the \textit{aliasing effect}).
It is useful at this stage to point out that the evolution of the soliton's
relative phase depends on their separation \cite{molecule1}. Therefore, each
result which we present is defined up to a choice of the initial phase
difference and separation, although altering these parameters does not yield
a qualitative difference.

\subsection{\label{Strong_Int}Strong interactions}

\begin{figure}[t]
\caption{(colour online). High-velocity collisions of two co-moving chiral
solitons. (a)-(b) Trajectories of inelastic interactions with $g_{\mathrm{1D}%
}m\ell /\hbar ^{2}=0$, $v_{1}m\ell /\hbar =2$, $v_{2}m\ell /\hbar =1$, and $%
\protect\delta =0$, where the gauge field strength is $a_{1}/\hbar =1$ in
(a), and $a_{1}/\hbar =4$ in (b). (c)-(d) Soliton fission with $g_{\mathrm{1D%
}}m\ell /\hbar ^{2}=2$, $v_{1}m\ell /\hbar =2$, $v_{2}m\ell /\hbar =0.5$, $%
a_{1}/\hbar =5$, with $\protect\delta =0$ (c), and $\protect\delta =\protect%
\pi $ (d). b.i) Populations of the soliton envelopes before and after the collision pictured in b), highlighting the population transfer.}
\label{solcol1}\includegraphics[width=9.3cm]{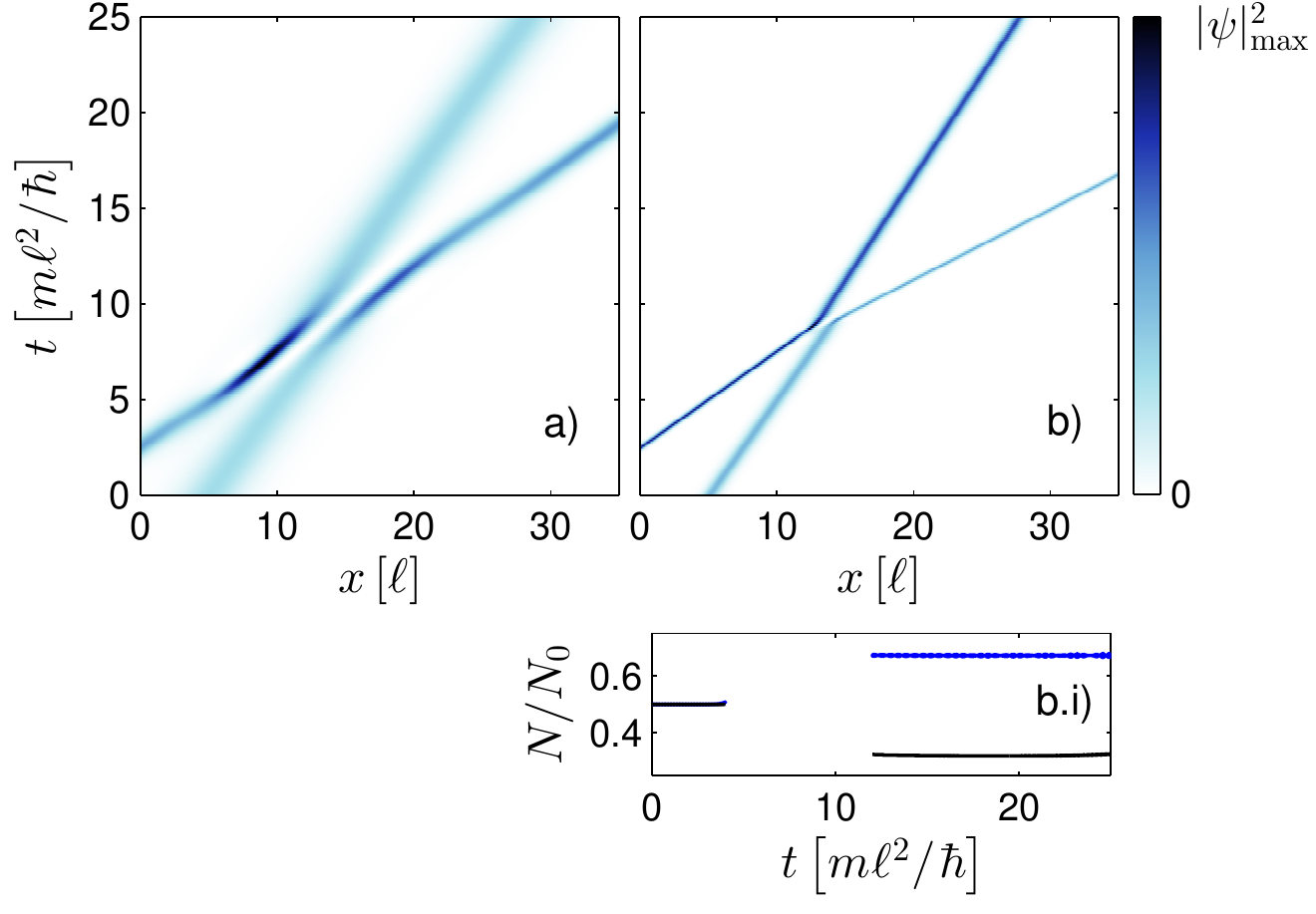} %
\includegraphics[width=9.3cm]{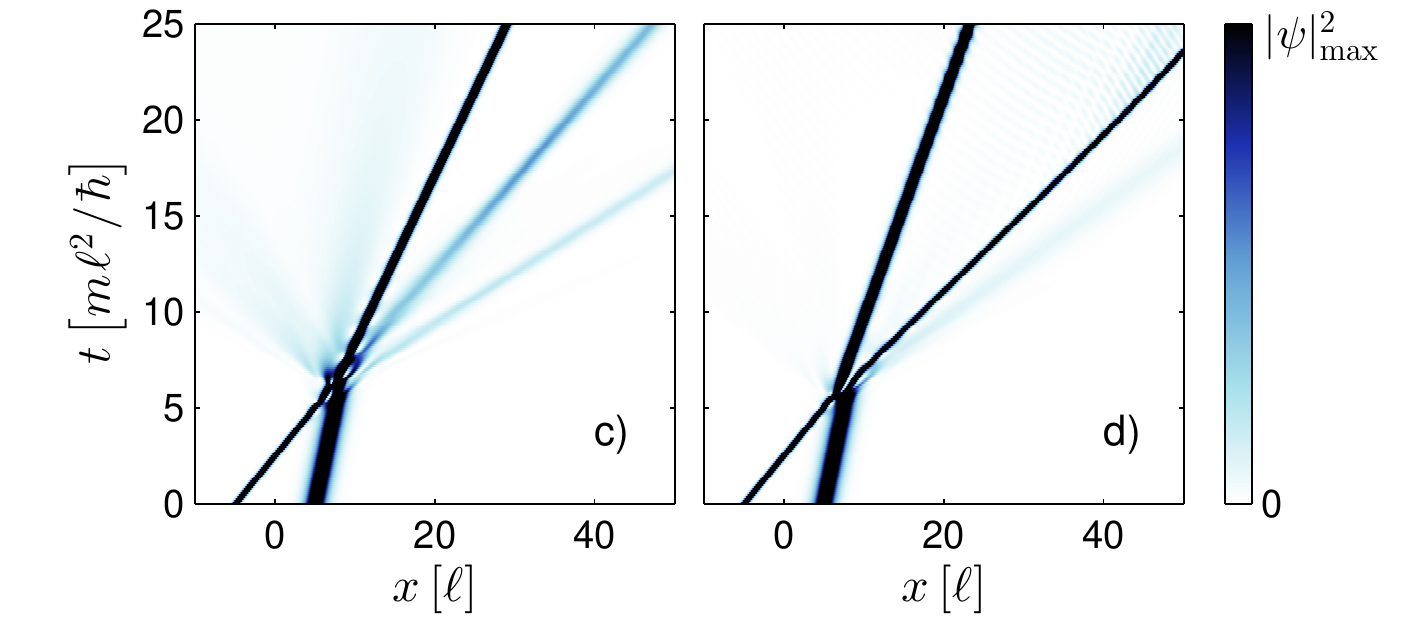} The colourbar
limit has been intentionally lowered in (c)-(d) to display the solitons more
clearly.
\end{figure}

In Fig. \ref{solcol1} we show a set of density plots for the collision of
two co-moving chiral solitons in the presence of the strong chiral
interactions.
In each case, the interaction induced by the current nonlinearity dominates
over the usual mean-field effects due to the solitons' high velocities,
which effectively reduces the mean-field scattering parameter. Figure~\ref%
{solcol1}(a) and Fig.~\ref{solcol1}(b) highlight extreme manifestations of
this regime, where the dynamics are solely influenced by the current
nonlinearity, setting $g_{\mathrm{1D}}m\ell /\hbar ^{2}=0$, i.e., the
mean-field contact interaction is completely disregarded.

Surprisingly, the collisions in these two instances are similar to those
produced by the conventional NLS dynamics, with the solitons surviving the
collision and retaining the general shape of their envelopes. However, two
key differences are visible in the trajectories of the solitons,
particularly in the case of $a_{1}/\hbar =4$, Fig.~\ref{solcol1}(b). The
first are a pair of inelastic trajectories, with velocities of the outgoing
solitons differing from their initial velocities, with the left- and
right-most solitons reducing and increasing their velocities, respectively.
The second difference from the mean-field setting is the appearance of a
density node at the interaction centre, which is reminiscent of a repulsive
interaction, despite the initial phase difference taken as $\delta =0$,
which is conventionally an attractive interaction (for the solitons
colliding in-phase). A third feature is also spotted in the form of population transfer, as illustrated in the Fig.~\ref{solcol1}(b.i) for the collision pictured in Fig.~\ref{solcol1}(b). Here, approximately a quarter of the initial mass of the right-most soliton is transferred to the left-most soliton in the course of the collision. Note, that the populations of the solitons are not calculated during the collision at $(\hbar/m\ell^{2})t=4$ to $(\hbar/m\ell^{2})t=12$, due to the overlap of the soliton envelops.

Each of these effects can be traced to the non-integrability of the current
nonlinearity, which permits the transfer of stored interaction energy into
the kinetic energy, and a possibility of a non-trivial shift of the soliton
phase difference. The presence of the population transfer is directly linked
to the phase shift, as both quantities are mutually conjugate. The energy
exchange, or, to a greater extent, the inelasticity of the collision,
appears to be minimised when the collision parameters are chosen so that the
solitons interact repulsively. This is evident from comparisons between the
two figures, where we note that $a_{1}/\hbar =1$ leads to a repulsive
interaction with elastic trajectories, while $a_{1}/\hbar =4$ gives rise to
a more attractive interaction which features inelastic trajectories.

The lower row of Fig.~\ref{solcol1} shows the dynamics where the collision
is destructive, causing fission of the solitons. In both cases, three
solitons emerge from the collision (the third soliton in Fig.~\ref{solcol1}%
(d) with $\delta =\pi $ is located at the leading edge of the other two)
with the populations and velocities of each outgoing soliton depending on
the initial phase difference. In addition, a modest amount of radiation is
ejected during the collision, as seen in the trailing edge in Fig.~\ref%
{solcol1}(c), and the interference pattern located between the slowest two
solitons in Fig.~\ref{solcol1}(d). The main difference against the previous
case is a larger difference in the initial velocities, which, if coupled
with a larger gauge-field strength, $a_{1}/\hbar =5$, produces two soliton
envelopes with a greater disparity of widths. As such, in the course of the
collision, the solitons effectively interact over a longer period, thus
enhancing effects stemming from the interaction.


\subsection{\label{Weak_Int}Weak interactions}

\begin{figure}[t]
\includegraphics[width=9.3cm]{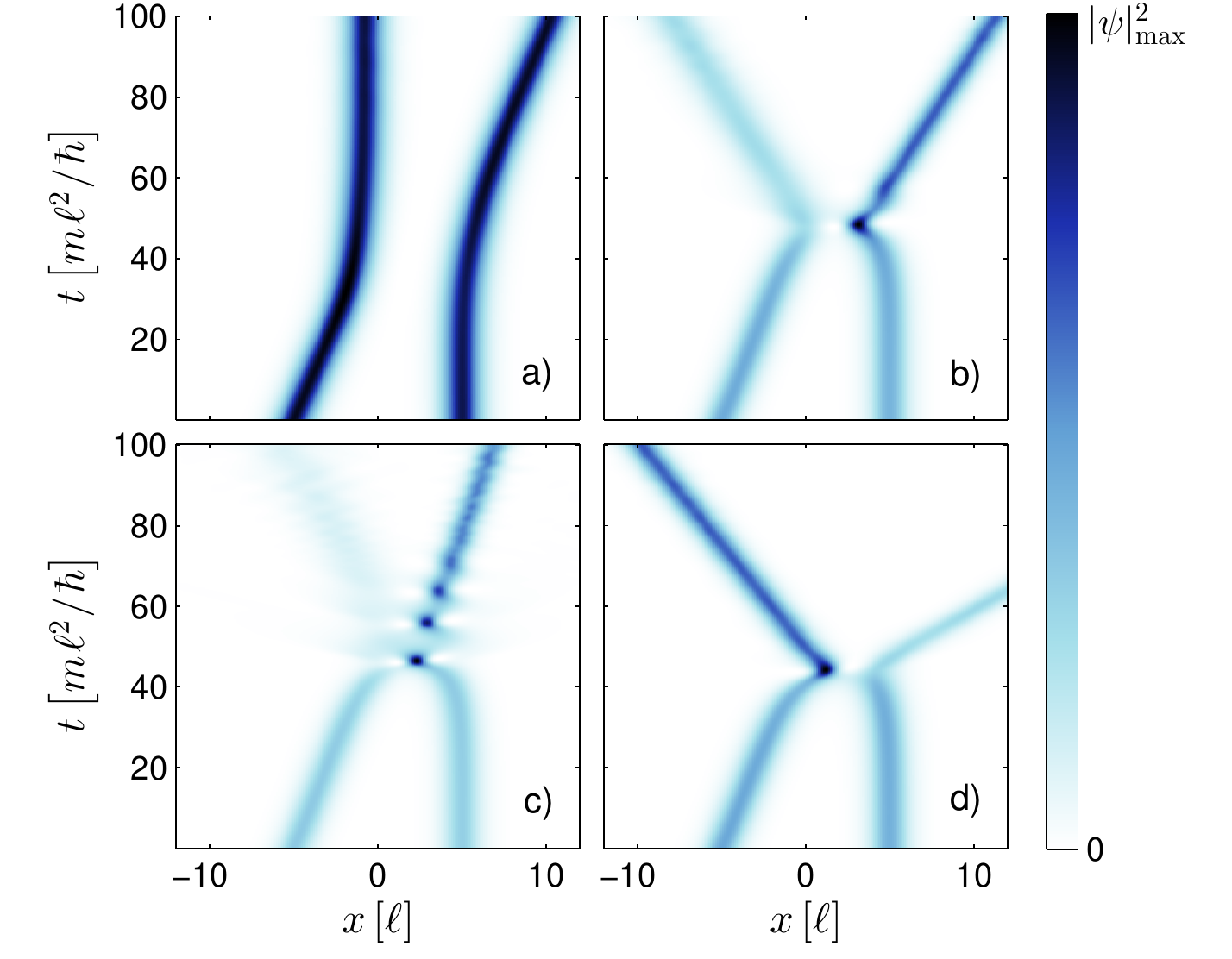}
\caption{(colour online). Asymmetric collisions between two chiral solitons
for various phase differences. The soliton parameters are $g_{\mathrm{1D}%
}m\ell /\hbar ^{2}=-4$, $v_{1}m\ell /\hbar =0.1$, $v_{2}m\ell /\hbar =0$,
and $a_{1}/\hbar =1.5$. The phase differences are taken as $\protect\delta =0
$ (a), $\protect\delta =0.9\protect\pi $ (b), $\protect\delta =0.98\protect%
\pi $ (c), and $\protect\delta =-0.9\protect\pi $ (d).}
\label{solcol2}
\end{figure}

In the previous subsection it was seen how chiral solitons undergo inelastic
collisions. To quantify the elasticity of the collisions, we introduce the
coefficient of restitution \cite{fusfis1}
\begin{equation}
\eta =\frac{\left( m_{1}v_{1}^{2}+m_{2}v_{2}^{2}\right) _{f}}{\left(
m_{1}v_{1}^{2}+m_{2}v_{2}^{2}\right) _{0}},  \label{COR}
\end{equation}%
which compares the difference of the kinetic energy before and after the
collision. For $\eta =1$, the collision is perfectly elastic with conserved
masses and velocities of the solitons, while $\eta \neq 0$ indicates an
inelastic collision. Here, $m_{1,2}$ and $v_{1,2}$ play the role of the
masses and velocities of the solitons in our semi-classical description, and
are calculated from the respective expectation values,
\begin{equation}
m_{1,2}=m\langle N_{1,2}\rangle =m\int dx|\psi |^{2},
\end{equation}%
\begin{equation}
v_{1,2}=\langle \hat{p}_{1,2}\rangle /m_{1,2}=-\frac{i\hbar }{m_{1,2}}\int
dx\psi ^{\ast }\partial _{x}\psi .
\end{equation}%
The integration in each case at either the initial or final time is
performed locally around each soliton's centre of mass to exclude
contributions from radiation and overlap with the other soliton. Due to the
occurrence of both the population transfer and changes in the outgoing
velocities of the solitons, one cannot distinguish whether the chiral
solitons pass through or rebound off each other during the collision.
Therefore, to remain consistent, we denote the solitons located in the $x<0$
and $x>0$ regions as the first and second ones, respectively.
\begin{figure}[t]
\includegraphics[width=9.25cm]{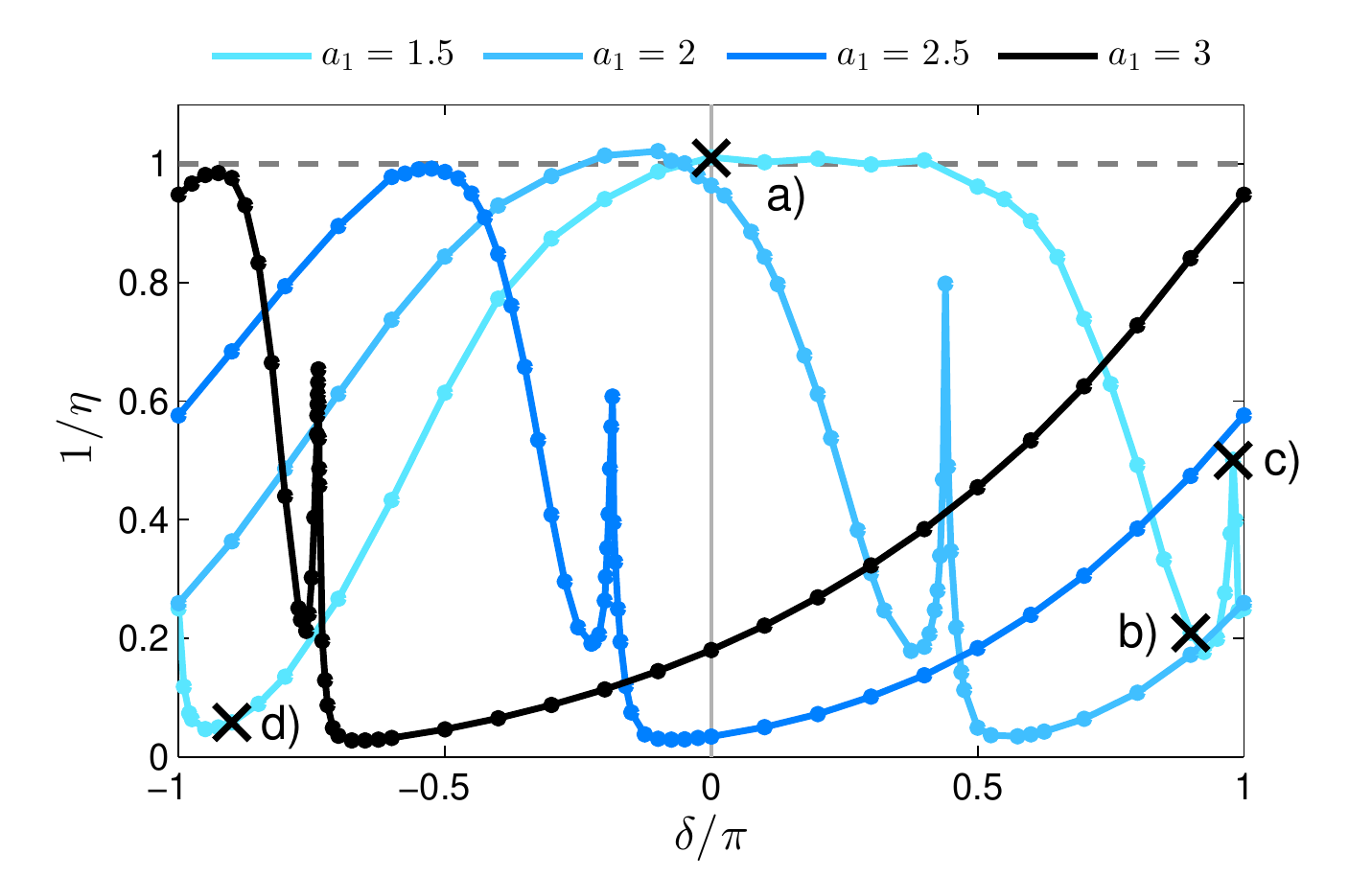}
\caption{(colour online). Inverse of the coefficient of restitution
calculated from numerical simulations (dots), with $g_{\mathrm{1D}}m\ell
/\hbar ^{2}=-4$, $v_{1}m\ell /\hbar =0.1$, and $v_{2}m\ell /\hbar =0$. The
grey dashed line indicates the standard GPE results with perfectly elastic
collisions, black crosses corresponding to the simulations in Fig. \protect
\ref{solcol2}.}
\label{solcol3}
\end{figure}
By varying the strength of the gauge field, we have performed a detailed
parameter scan of the soliton-soliton collisions as a function of the
initial phase difference, with the initial soliton velocities fixed. The
coefficient of restitution is computed and plotted in Fig.~\ref{solcol3}
with corresponding examples of the dynamics shown in Fig.~\ref{solcol2}.
For each value of the gauge-field strength, three regimes of the collision
dynamics can be identified, depending on the initial phase difference
between the solitons. The first is an elastic scattering regime highlighted
by a plateau in the restitution data at $\eta =1$, with an example of the
dynamics shown in Fig.~\ref{solcol2}(a). Here, the interaction is notably
repulsive, with a distinct node in the density at the interaction centre and
the soliton parameters keeping their values after the collision.

Away from this plateau, two distinct regimes of inelastic dynamics are found
with $\eta >1$, as illustrated in Fig.~\ref{solcol2}(b) and Fig.~\ref%
{solcol2}(d). Here, the dynamics are also similar to the case of strong
interactions, with inelastic trajectories that feature a redistribution of
the soliton masses, as well as evolution of the initial phase difference,
resulting in shifts of the in- and out-of-phase collision points. Comparing
these two plots, one notices that, depending on the direction in which one
moves away from the plateau in the parameter space, the soliton mass can be
transferred, chiefly, in either the left (Fig.~\ref{solcol2}(d)) or
right-hand (Fig.~\ref{solcol2}(b)) outgoing soliton.

The final inelastic regime, indicated by the `resonance' peak in the
restitution data represented by the cross labelled (c) in Fig.~\ref{solcol3}%
), features the turning point of this population transfer, where the mass is
transferred from one soliton to the other. We show in Fig.~\ref{solcol2}(c)
an example of the dynamics in this regime, at the maxima of the `resonance'
peak. Here, a peculiar soliton state is formed, where there is a strong
interplay between emitted radiation, excitation of an oscillatory mode in
the right-hand soliton, and a weak left-hand one. This regime appears to be
an example of in-phase (fully attractive) dynamics, which features the
formation of a metastable (short-lived) bound state.

A feature universal to the restitution data presented in Fig.~\ref{solcol3}
is that the location of each inelastic regime is cyclically shifted
left-wards for an increasing current strength. Comparing different
gauge-potential strengths, one can see that the elastic region shrinks for
larger values, which can be explained by enhancement of the
non-integrability effects for stronger gauge-field strengths. Although not
shown here, the dip in the restitution data initially appears close to $%
\delta =0$ at small values of the current strength, and cyclically displaces
towards lower $\delta $ for increasing current strengths.

\begin{figure}[t]
\includegraphics[width=9.3cm]{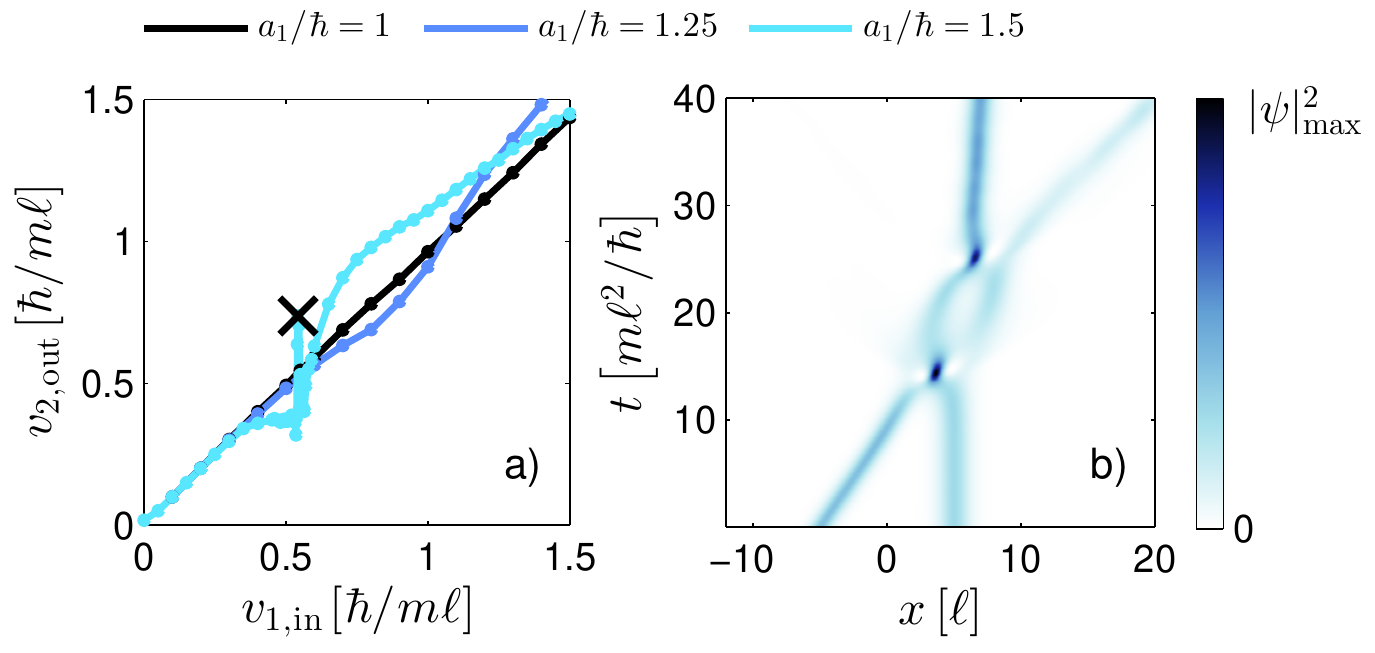}
\caption{(colour online). (a) Outgoing versus incoming velocities for
asymmetric collisions between two chiral solitons for various gauge-field
strength, with parameters $g_{\mathrm{1D}}m\ell /\hbar ^{2}=-4$, $v_{2}m\ell
/\hbar =0$, and $\protect\delta =0$ fixed in each instance. Black cross
indicates the two-bounce resonance state shown in (b) for $v_{1}m\ell /\hbar
=0.5425$ and $a_{1}/\hbar =1.5$.}
\label{solcol5}
\end{figure}

To complete the analysis for the weak-chiral regime, we perform a similar
parameter scan as before, but now in the case when the relative phase
difference is fixed to $\delta =0$, with the initial velocity of the
left-hand soliton allowed to vary. In this case, the coefficient of
restitution provides a poor illustration of the underlying dynamics,
therefore we, instead, plot the outgoing velocity of the soliton travelling
to the right for increasing values of the gauge-field strength, as shown in
Fig. \ref{solcol5}(a).

Depending on the choice of the initial velocity and gauge-field strength,
the strength of the chiral interactions $|a_{1}(v_{1}+v_{2})|$ may be either
small or comparable to the mean-field strength $|g_{\mathrm{1D}}|$.
Therefore, for extreme values of the parameters, it is expected that the
dynamics will be generally inelastic in a similar manner to Fig. (\ref%
{solcol1}), whereas for smaller values the dynamics will be, generally,
elastic. This reasoning is reflected in the pair of curves corresponding to $%
a_{1}/\hbar =1$ and $a_{1}/\hbar =1.25$ in Fig. \ref{solcol5}(a), in which
the soliton velocity does not change significantly after the interaction for
small initial velocities. As the velocity increases (and hence the
interaction strength increases too), this invariance begins to break, which
is particularly notable in the case of $a_{1}/\hbar =1.25$, which exhibits a
sinusoidal behaviour, at the velocity exceeding a critical value, $%
v_{1}\approx 0.5$.

As the gauge field strength is increased further, as in the case of $%
a_{1}/\hbar =1.5$, a `resonance' feature appears in the data where a
two-bound resonance state is formed, as shown in Fig. \ref{solcol5}(b). As
mentioned previously, such states are a common occurrence in non-integrable
models \cite%
{chaos1,chaos2,chaos4,chaos5,chaos6,chaos7,chaos8,chaos9,chaos10,chaos11},
perhaps most notably in non-integrable versions of the sine-Gordon equation
\cite{chaos11}, where, depending on the strength and shape of the
interaction potential, an $n$-bound resonance state may emerge. Here the
underlying mechanism is the energy exchange between the colliding solitons
as a whole and their internal modes, which requires the solitons to collide
several times before escaping, thus regaining the energy temporarily
transferred into the internal mode. In performing this parameter scan,
higher-order bound states, where the solitons collide more than twice, were
not observed, as for stronger interaction strengths the appearance of a
bound state tends to be suppressed in a similar manner to Fig. \ref{solcol2}%
(c).

\subsection{\label{Bound_states}Bound states}

\begin{figure}[t]
\includegraphics[width=9.3cm]{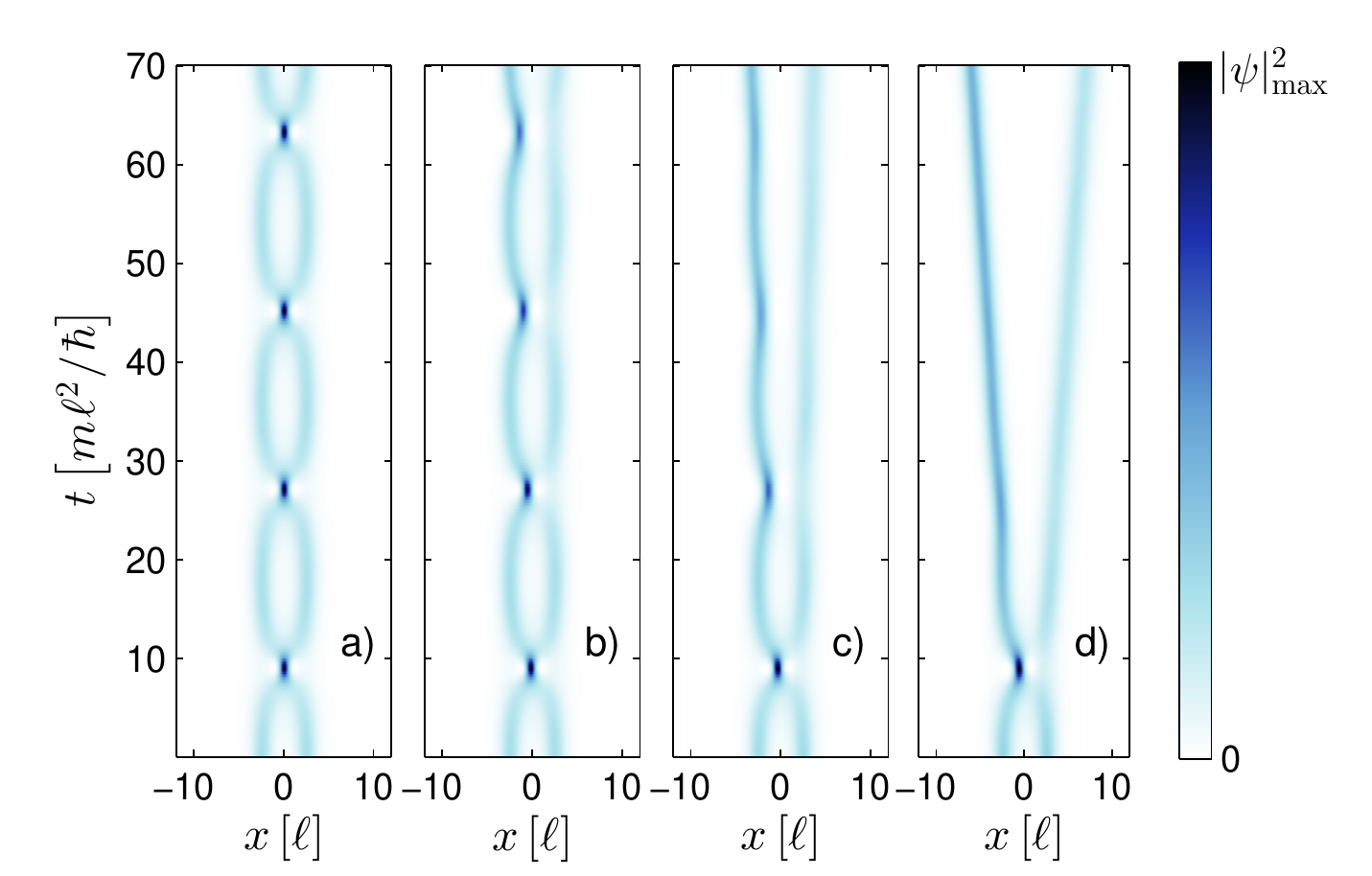}
\caption{(colour online). Breakdown of the soliton-soliton bound state due
to the presence of the current nonlinearity. Two stationary solitons are
initially placed at distance $x/\ell =5$ units apart, with $g_{\mathrm{1D}%
}m\ell /\hbar ^{2}=-4$ and $\protect\delta =0$. The gauge-field strength
varies as $a_{1}/\hbar =0$ (a), $a_{1}/\hbar =0.125$ (b), $a_{1}/\hbar =0.25$
(c), and $a_{1}/\hbar =0.5$ (d).}
\label{solcol4}
\end{figure}

To further investigate the inelastic dynamics of the density-dependent gauge
theory, we consider a set of symmetric collisions (see Fig. \ref{solcol4}),
where two interacting solitons form a molecule-like bound state.
In a similar manner to results obtained in weakly perturbed cubic-quintic
NLS \cite{chaos1,chaos10} and sine-Gordon systems \cite{chaos2}, a weak
current nonlinearity is found to support short-lived bound states, where the
solitons collide several times before escaping. As before, the underlying
mechanism here is the transfer of a part of the energy of the interaction
between the solitons into the kinetic energy and redistribution of the
soliton masses. However, the magnitude of the interaction energy is now
comparable to or larger than the kinetic energy of the solitons, requiring
them to collide several times in order to gain enough kinetic energy for
escaping the attractive interaction. Compared to the standard
Gross-Pitaevskii dynamics shown in Fig.~\ref{solcol4}(a) with $a_{1}/\hbar =0
$, where the solitons are perpetually trapped with a fixed oscillation
amplitude and frequency, a modest current strength can begin to destabilize
the bound state, such as in Fig.~\ref{solcol4}(c) for $a_{1}/\hbar =0.25$,
where the solitons collide four times before escaping, and in Fig.~\ref%
{solcol4}(d) for $a_{1}/\hbar =0.5$, where they collide twice before
escaping.

Interestingly, despite the interaction being initially symmetric, effects
stemming from the chiral dynamics result in a left-handedness in the
post-collision behaviour. For example, in Fig.~\ref{solcol4}(b) with $%
a_{1}/\hbar =0.125$, the first collision at $(\hbar /m\ell ^{2})t=10$ is
noticeably attractive due to the presence of the anti-node at the
interaction centre, but every subsequent collision becomes increasingly
repulsive with the amplitude of the anti-node decreasing and its position
shifting towards the left. In addition, a density node fills the vacancy
left by the anti-node at each interaction centre, with some manifestation of
the population transfer. This effect is seen to be most profound in Fig.~\ref%
{solcol4}(d) for $a_{1}/\hbar =0.5$, where $\sim 60\%$ of the outgoing mass
in captured in the left soliton. Returning to the two-bound resonance state
in Fig. \ref{solcol5}, the existence of higher-order bound states appears
unlikely due to the fact that each subsequent interaction becomes more
repulsive, hence more elastic than the previous one. These dynamics
highlight the role that the lack of the Galilean invariance plays in the
interacting gauge theory. In particular, the current operator appearing in
Eq.~\eqref{gaugeEOM} induces the population transfer between the two
solitons, resulting in the suppression and, ultimately, breakdown of the
bound state.

\section{\label{Variational}The variational analysis}

To gain insight into how the current nonlinearity modifies the interactions
between the solitons, we have performed variational calculations to derive
an effective particle model for the soliton dynamics. We achieve this by
using two similar, but essentially different methods.

In the first instance, we approximate the two-soliton state as a linear
superposition of two individual solitons, with the interaction treated as
the spatial overlap of the soliton envelopes. This technique has been
previously applied to interaction problems in the NLS \cite{solint2,solint3}
and Gross-Pitaevskii equations \cite{solint4,solint6}, in addition to
several others \cite{solint1,solint5,solint7,solint8,solint9}. An advantage
of this method is the ability to derive a set of variational equations which
describe the motion of the solitons. From this, two key results can be
extracted. The first is an effective potential describing the interaction
between the solitons, which will provide details into the phase dependence
and range of the interactions. Secondly, by numerically solving the
variational equations, we will be able to illustrate the dynamics of the
particle model and compare it directly to the full numerical solutions that
are presented above.

For the second method, we follow the technique outlined in Refs. \cite%
{boris1,boris2}, in which the soliton state is also approximated as a linear
superposition, but restricted to the case of two stationary solitons which
are well-separated. In this case, the interaction is accounted for by the
spatial overlap of one soliton with the `weak tail' of the other, and may
therefore be regarded as an asymptotic approximation to the full
interaction. As such, this method is well suited to the study of bound
states, but will also provide a basis to draw comparisons to the first
method in a far-field low-velocity limit.

\subsection{\label{sec:level2}Linear calculations}

The starting point for the variational calculation is the Lagrangian density
\cite{anyon,zoller}
\begin{equation}
\begin{split}
\mathcal{L}& =\frac{i}{2}\left( \tilde{\psi}\partial _{\tilde{t}}\tilde{\psi}%
^{\ast }-\tilde{\psi}^{\ast }\partial _{\tilde{t}}\tilde{\psi}\right) +\frac{%
1}{2}|\partial _{\tilde{x}}\tilde{\psi}|^{2}+\frac{\tilde{g}_{\mathrm{1D}}}{2%
}|\tilde{\psi}|^{4} \\
& +\tilde{a}_{1}|\tilde{\psi}|^{2}\frac{d}{d\tilde{t}}\int_{-\infty }^{\tilde{x}}d%
\tilde{x}^{\prime }|\tilde{\psi}\left( \tilde{x}^{\prime },\tilde{t}\right) |^{2},
\end{split}%
,  \label{Lagrangian}
\end{equation}%
where we have introduced dimensionless variables $x=\tilde{x}\ell $, $t=%
\tilde{t}m\ell ^{2}/\hbar $, $\psi =\tilde{\psi}/\sqrt{\ell }$, $g_{\mathrm{%
1D}}=\tilde{g}_{\mathrm{1D}}\hbar ^{2}/m\ell $, and $a_{1}=\tilde{a}%
_{1}\hbar $, with the length scale $\ell $. Here, and in the remainder of
this paper, we drop the tildes appearing in these variables, working in the
dimensionless units.

In what follows, we seek stationary solutions determined by the action
\begin{equation}
\begin{split}
\mathcal{S}\left[ q_{n}(t)\right] & =\int_{t_{1}}^{t_{2}}dt\int_{-\infty
}^{\infty }dx\mathcal{L}(q_{n}(t),\dot{q_{n}}(t)) \\
& =\int_{t_{1}}^{t_{2}}dt\langle \mathcal{L}\rangle ,
\end{split}
\label{action}
\end{equation}%
which gives rise to the Euler-Lagrange equations for each variational
parameter $q_{n}(t)$, namely,
\begin{equation}
\frac{\partial \langle \mathcal{L}\rangle }{\partial q_{n}}-\frac{d}{dt}%
\frac{\partial \langle \mathcal{L}\rangle }{\partial \dot{q_{n}}}=0.
\label{EL}
\end{equation}%
The variational ansatz is adopted as
\begin{equation}
\psi =\sum_{j=1,2}a\sech\left( (x-\xi _{j})/b\right) e^{iS_{j}},
\label{Solansatz}
\end{equation}%
in which each soliton contains a spatially-varying phase \cite%
{solansatz1,solansatz2},
\begin{equation}
S_{j}=v_{j}\left( x-\xi _{j}\right) +\phi _{j}.  \label{Solphase}
\end{equation}%
Here, $a(t)$, $b(t)$, $\xi _{j}(t)$, $v_{j}(t)$, and $\phi _{j}(t)$ are
time-dependent variational parameters corresponding to the amplitude, width,
centre-of-mass coordinates, velocities, and central phases of the solitons.
This ansatz models two bright solitons in which the individual velocities
and positions are allowed to evolve independently, with the interaction
treated as the (linear)-overlap of the soliton envelopes. The constraint
that the solitons have a common width, which in turn fixes the profiles of
the soliton envelopes, is a necessary restriction in order to be able to
explicitly calculate interaction integrals. Consequently, this restricts the
variational analysis to the regime in which
\begin{equation}
\frac{b_{1}}{b_{2}}=\frac{g_{\mathrm{1D}}-2a_{1}v_{2}}{g_{\mathrm{1D}%
}-2a_{1}v_{1}}\approx 1.
\end{equation}%
This can be achieved by considering collisions with small velocities and by
compensating the effects of the gauge field by a mean field with a modest
strength, such that $|a_{1}(v_{1}+v_{2})|\ll |g|$. Therefore, the above
constraint restricts our variational analysis to the weak-chiral regime for
which numerical results are presented above. In spite of these restrictions,
we will find that one may be quite liberal with the choice of parameters and
still achieve sensible results.

Our choice of ansatz arises due to two reasons. First, our model is
non-integrable, therefore a closed-formed expression for a two-soliton state
via inverse scattering techniques is not available. Secondly, regardless of
whether such a solution existed, Eq. (\ref{Solansatz}) should work as a good
approximation to the dynamics pictured in Fig. (\ref{solcol2}), as the
solitons roughly retain their shape during the interaction. However, it must
be stressed that this choice of the ansatz does not fully replicate all the
features of the interaction and will therefore lead to inconsistencies at
short length scales, when the solitons begin to significantly overlap.

An important example of this which must be considered before proceeding, is
related to a divergence of the soliton amplitude $a(t)$ at short length
scales. This can be illustrated by evaluating Eq. (\ref{density}) for our
variational ansatz, from which one can obtain the following expression for
the soliton's amplitude:
\begin{equation}
a^{2}=\frac{1}{4b}\frac{1}{1+\epsilon \csch(\epsilon )\cos (\delta )}=\frac{1%
}{4b}\frac{1}{f(\epsilon ,\delta )}.  \label{norm}
\end{equation}%
To perform the integration, we have introduced the change of variables $%
\alpha =(x-\xi _{1})/b$ and $\alpha +\epsilon =(x-\xi _{2})/b$, where $%
\epsilon =(\xi _{1}-\xi _{2})/b$ is a new variational parameter describing
the relative positions of the solitons. Additionally, we have assumed that
the magnitude of the velocity is small, such that the phase difference $%
S_{1}-S_{2}\approx \delta $ is an approximate function of solely the central
phases. Additional details of the calculation are outlined in Appendix. \ref%
{integrals}. Although this integral can be evaluated exactly without needing
this approximation \cite{solint3}, the ensuing exact expression is to
cumbersome for extracting explicit results from it. Inspecting Eq. (\ref%
{norm}), one can see that, in the limit of $\epsilon \rightarrow 0$, the
value of $a^{2}(t)$ rapidly diverges for $\delta \in \lbrack \pi /2,3\pi /4]$
and approaches a singularity at $\delta =\pi $, modulo $2\pi $. Our
approximations therefore lead to an unphysical divergence which is not
representative of soliton collisions, thereby requiring us to restrict our
studies to the bounded domain $|f(\epsilon ,\delta )|\geq 1$, which
corresponds to the interval of $\delta \in \lbrack 0,\pi /2]$. For such
values of $f$ which are bounded from below, $a(t)$ is non-divergent,
representing the known dynamics more adequately---for instance, at $\delta =0
$, which corresponds to the fact that the solitons' amplitude is increased
by $1/\sqrt{2b}$ when they constructively interfere.

Substituting our ansatz into Eq. (\ref{Lagrangian}) and integrating via Eq. (%
\ref{action}) leads to the averaged Lagrangians
\begin{widetext}
\begin{equation}\label{Leff0}
\langle\Lagr_0\rangle=2a^2b(\dot{\phi}_1+\dot{\phi}_2)+\frac{4a^4bg^\prime_2}{3}+\frac{2a^2}{3b}-2a^2b(v_1\dot{\xi}_1+v_2\dot{\xi}_2)+a^2 b(v_1^2+v_2^2),
\end{equation}
and
\begin{equation}\label{LeffI}
\begin{split}
\langle\Lagr_I\rangle&=2a^4bg^\prime_2\left(1+\cos(2\delta)/2\right)\left[\frac{4\epsilon\cosh(\epsilon)}{\sinh^3(\epsilon)}-\frac{4}{\sinh^2(\epsilon)}\right]+\frac{a^2}{b}\cos(\delta)\left[\frac{4\cosh(\epsilon)}{\sinh^2(\epsilon)}+\frac{4\epsilon}{\sinh(\epsilon)}-\frac{4\epsilon\cosh^2(\epsilon)}{\sinh^3(\epsilon)}\right]\\
&+a^2(v_1+v_2)\sin(\delta)\left[\frac{2\epsilon\cosh(\epsilon)}{\sinh^2(\epsilon)}-\frac{2}{\sinh(\epsilon)}\right]-a^2b^2\cos(\delta)\left(\dot{v}_1-\dot{v}_2\right)\left[\frac{\epsilon^2}{\sinh(\epsilon)}\right]\\
&+a^2b\cos(\delta)\left(v_1v_2-(v_1\dot{\xi}_1+v_2\dot{\xi}_2)+\dot{\phi}_1+\dot{\phi}_2\right)\left[\frac{2\epsilon}{\sinh(\epsilon)}\right]+4a^4bg^\prime_2\cos(\delta)\left[\frac{2\cosh(\epsilon)}{\sinh^2(\epsilon)}-\frac{2\epsilon}{\sinh^3(\epsilon)}\right],
\end{split}
\end{equation}
\end{widetext}where we have defined $g_{2}^{\prime }=g_{\mathrm{1D}%
}-a_{1}\left( v_{1}+v_{2}\right) $, and have split the total Lagrangian into
the sum of terms which implicitly and explicitly depend on $\epsilon $, as
denoted by the free and interacting Lagrangians, $\mathcal{L}=\mathcal{L}%
_{0}+\mathcal{L}_{I}$.

Equations of motion for each variational parameter can now be derived from
Eq. (\ref{EL}), which lead to the set of coupled differential equations
\begin{widetext}
\begin{equation}\label{phi}
\phi_i:\quad\quad\partial_t\left(4a^2bf\left(\epsilon,\delta\right)\right)=0,
\end{equation}
\begin{equation}\label{v1}
v_{i=1(+), 2(-)}:\quad\quad v_i=\dot{\xi}_il_i\left(\epsilon,\delta\right)+\frac{4a^2a_1}{6}+d\left(\epsilon,\delta\right)-v_{i\pm 1}\cos(\delta)\frac{\epsilon}{\sinh(\epsilon)},
\end{equation}
\begin{equation}\label{a}
a:\quad\quad\left.\frac{\partial \langle\Lagr\rangle}{\partial a}\right|_{\dot{q}_n}+4ab\sum_{n=1,2}\left(\dot{\phi}_n-v_n\dot{\xi}_n\right)f(\epsilon,\delta)=2ab^2\left(\dot{v}_1-\dot{v}_2\right)\cos(\delta)\frac{\epsilon^2}{\sinh(\epsilon)},
\end{equation}
\begin{equation}\label{xi1}
\xi_i:\quad\quad\left.\frac{\partial \langle\Lagr_I\rangle}{\partial \xi_i}\right|_{\dot{q}_n}+2a^2b\frac{\partial f}{\partial\xi_i}\sum_{n=1,2}\left(\dot{\phi}_n-v_n\dot{\xi}_n\right)=-2\partial_t\left( a^2bv_if(\epsilon,\delta)\right)+a^2b^2\left(\dot{v}_1-\dot{v}_2\right)\partial_{\xi_i}\frac{\epsilon^2}{\sinh(\epsilon)},
\end{equation}
with
\begin{equation}\label{d-function}
\begin{split}
d\left(\epsilon,\delta\right)&=4a^2a_1\left(\left(1+\frac{\cos(2\delta)}{2}\right)\left[\frac{\epsilon\cosh(\epsilon)}{\sinh^3(\epsilon)}-\frac{1}{\sinh^2(\epsilon)}\right]+\cos(\delta)\left[\frac{\cosh(\epsilon)}{\sinh^2(\epsilon)}-\frac{\epsilon}{\sinh^3(\epsilon)}\right]\right)\\
&-\frac{\sin(\delta)}{b}\left[\frac{\epsilon\cosh(\epsilon)}{\sinh^2(\epsilon)}-\frac{1}{\sinh(\epsilon)}\right],
\end{split}
\end{equation}
and
\begin{equation}\label{l-function}
l\left(\epsilon,\delta\right)_{i=1(+), 2(-)}=1+\cos(\delta)\frac{\epsilon}{\sinh(\epsilon)}\mp\frac{1}{2a^2b}{\partial}_{\xi_i}\left(a^2b^2\cos(\delta)\frac{\epsilon^2}{\sinh(\epsilon)}\right).
\end{equation}
\end{widetext}Here, the vertical bar notation in Eqs. (\ref{a}-\ref{xi1}),
denotes the full Lagrangian function in Eq. (\ref{LeffI}), but excluding
terms containing a factor of $\dot{\phi}_{i}$, $\dot{\xi}_{i}$, or $\dot{v}%
_{i}$. In Eqs. (\ref{v1}) and (\ref{l-function}), for $i=1$ positive (+)
operations are taken, with the converse for $i=2$. A variational equation
for $b$, the solitons' width, is not required to proceed and is excluded.
Both Eqs. (\ref{d-function}) and (\ref{l-function}) are introduced for
notational convenience. From the set of variational equations, we can now
extract details of how the gauge field affects the soliton dynamics, {and
derive several important quantities. }

Starting with the first variational equation, which can be obtained by
varying either $\phi _{1}$ or $\phi _{2}$, one can identify Eq. (\ref{phi})
as a conservation law for $4a^{2}bf(\epsilon ,\delta )$. This is consistent
with both Eqs. (\ref{density}) and (\ref{norm}), which state that the phase
and density of the condensate are conjugate variables. In the asymptotic
limit of $\epsilon \rightarrow \infty $, this conserved quantity reduces to $%
a^{2}\sim 1/4b$, which is the correct amplitude for a two-soliton state.

The equations for the velocities $v_{i}$ highlight the main result of the
variational analysis. The first and last terms of Eq. (\ref{v1}) (and the
last term of Eq. (\ref{d-function})), correspond to similar terms in the NLS
equation. Together they imply that, in the asymptotic limit of $v_{i}\sim
\dot{\xi _{i}}$, both solitons move at a constant velocity when they are
well separated. However, at $\epsilon \rightarrow 0$ the velocities of the
solitons are modified due to their interaction, but, once again, they become
constant after the solitons have passed through each other. The additional
terms $\propto a_{1}$, which appear in the velocity equation, are new ones,
which arise due to the presence of the gauge potential. The first of these
is a non-Galilean effect that redefines the soliton velocities in the
asymptotic limit as $v_{i}\sim \dot{\xi}+a_{1}/6b$, which is consistent with
the momentum conservation law stated in Eq. (\ref{momentum}). The remaining
terms which appear in Eq. (\ref{d-function}) are responsible for the
interaction-induced velocity shift, which, in both the $v_{1}$ and $v_{2}$
equations, has the same magnitude and sign.

The variational equations for both $a$ and $\xi _{i}$ are not particularly
transparent. However, they do highlight the coupling between all of the
variational parameters, and will be required when deriving the interaction
potentials later in the section.

\subsubsection{\label{sec:level3}Collision dynamics}

In order to illustrate how the gauge field is the mechanism underlying the
inelastic scattering in our system, we set out to first simplify and reduce
the number of variational equations, so that an effective particle model can
be derived. Subsequently, we can numerically solve our system of equations
and compare it to the full numerics presented above.

We begin by first reiterating that we consider the case of weak-chiral
interactions $|a_{1}(v_{1}+v_{2})|\ll |g|$, which feature the solitons
moving slowly for a given choice of the gauge-field strength. Equations (\ref%
{a}-\ref{xi1}) may then be set up as a set of simultaneous equations in
which coordinates, $\dot{\phi}_{1}$, $\dot{\phi}_{2}$, $\dot{%
\xi}_{1}$, and $\dot{\xi}_{2}$ can be eliminated, leading to a pair of
equations
\begin{equation}
\dot{v}_{1}\gamma +\dot{v}_{2}\left( \gamma -1\right) =-4\left. \frac{%
\partial \langle \mathcal{L}_{I}\rangle }{\partial \xi _{1}}\right\vert _{%
\dot{q}_{n}},  \label{difxi1}
\end{equation}
\begin{equation}
\dot{v}_{2}\gamma +\dot{v}_{1}\left( \gamma -1\right) =-4\left. \frac{%
\partial \langle \mathcal{L}_{I}\rangle }{\partial \xi _{2}}\right\vert _{%
\dot{q}_{n}},  \label{difxi2}
\end{equation}%
with $\gamma \equiv 1-(b/2)\partial _{\xi _{1}}\left( \epsilon \left(
f-1\right) /f\right) $. Together with Eq. (\ref{v1}), they form a set of
coupled differential equations for the soliton dynamics, in which details of
the interaction are encoded in expressions for $\gamma (\epsilon ,\delta )$
and $\langle \mathcal{L}_{I}\rangle $. Once again, in the asymptotic limit,
both equations simplify to the single-soliton result $\dot{v}_{i}\sim 0$,
which highlights that both solitons move independently at a constant
velocity when they are well separated. A consequence stemming from the
elimination of variables $\dot{\phi _{1}}$ and $\dot{\phi _{2}}$ in the
equations highlights that the phase in this particle model is \textit{static}%
, and does not dynamically evolve. Although this is an important feature in
our model which has many consequences in the scattering dynamics, we will
still be able to obtain qualitative results which do not strongly depend on
phase $\delta $, but will not be able to address issues pertaining to the
bound-states dynamics pictured in Fig. \ref{solcol4}.

\begin{figure}[t]
\includegraphics[width=7.0cm]{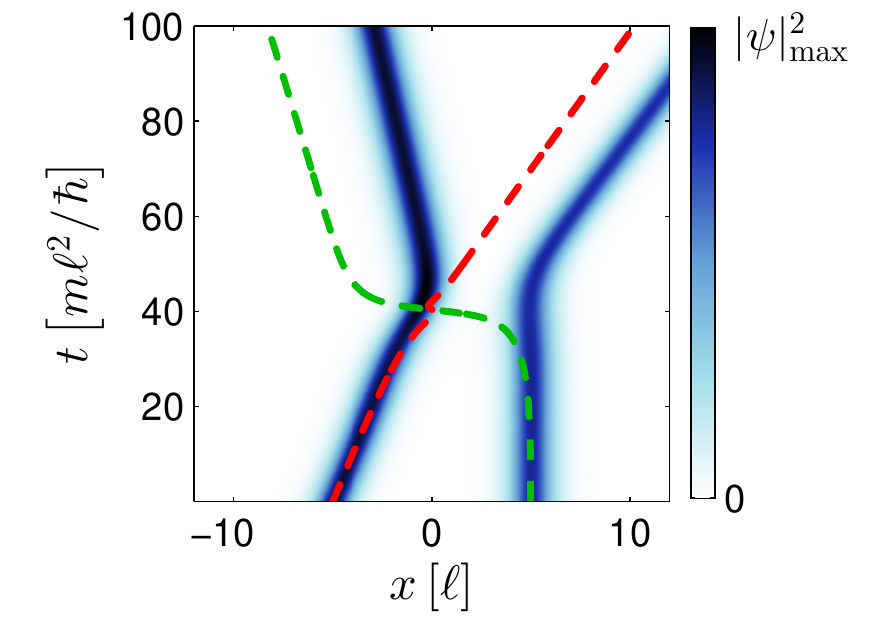}
\caption{(colour online). Comparison between the solutions of the
variational equations (dashed-green/red) and full numerics (blue) for the
evolution of the solitons' centres of mass. The soliton parameters are taken
as $g_{\mathrm{1D}}=-4$, $v_{1}=0.1$, $v_{2}=0$, $a_{1}=3$, and $\protect%
\delta =0.4\protect\pi $. }
\label{Rungepic}
\end{figure}

We solve the set of differential equations numerically using a fourth-order
Runge-Kutta method and compare our results to an example of the full
numerics in Fig. (\ref{Rungepic}).
For the chosen set of parameters, the magnitude of the outgoing velocities
are in good agreement with the post-collision trajectories showing that the
solitons pass through each other. However, position shifts of the solitons
are not captured well, with the left-outgoing soliton shifted too much, and
the right-outgoing soliton shifted too little. This particular example
represents the configuration that has the best agreement for the velocities.
Although not shown here, for $\delta <0.3\pi $ the particle model predicts
that the solitons form a perpetual bound state with a centre of mass
coordinate that increases linearly with time. Otherwise, for $\delta >\pi /2$%
, the dynamics feature a hard-core elastic interaction where the solitons
collide, but rebound off each other. Although the dynamics in these two
regimes are similar to what we have obtained numerically, in that we can
identify regimes where the interaction is repulsive (and therefore elastic)
and attractive (supporting bound states), this correlation actually arises
from \textit{discrepancies} in our model.

To explain how these discrepancies appear in our results, we restate the
consequences of the various approximations that we have used in the
analysis. In effect, all inconsistencies can be traced back to the initial
ansatz used in the analysis, see Eq. (\ref{Solansatz}). The first problem is
the obvious fact that the ansatz is a linear superposition of two solitons,
neglecting the nonlinear deformation which takes place when they overlap
significantly. For this reason, important details of the interaction are
omitted.

The second discrepancy in the ansatz, arises from the need to fix and equate
the soliton widths. For two chiral solitons to interact, their velocities,
and therefore, by extension, their widths, must be different. Furthermore,
as the solitons' velocities can change after colliding, $b(t)$ is a
time-dependent quantity that requires the additional term $w\left( x-\xi
_{i}\right) ^{2}$ in the expression for the phase given by Eq. (\ref%
{Solphase}), with $w$ being the \textit{chirp}. Although this was derived,
it was eventually excluded due the complexity in implementing it in the
particle model. The net result is that interaction effects in our particle
model are isotropic with respect to each soliton's mutual influence, which
is clearly not the case in the full numerics.

Another inconsistency is the divergence of the amplitude Eq. (\ref{norm})
for short length scales in the regime of $\delta >\pi /2$. This artifact
enters due to approximating the phase difference as a function of only the
central phases, therefore neglecting velocity contributions. Although this
was justified by taking the velocities small, the spatially-varying form of
this phase is required to obtain sensible results, as was shown in \cite%
{solint3}. Due to the divergence of the amplitude, the effective interaction
potential between the solitons also diverges in this regime, as we will
demonstrate in the following subsection.

The final discrepancy takes place due to the static nature of the phase in
our model. As was demonstrated in our simulations, the current nonlinearity
introduces the population transfer and shifts in the soliton's central phase
at each collision. The absence of these properties in the model results,
therefore, in the existence of perpetual bound states, in addition to the
lack of changes in the soliton's amplitudes/widths due to the populations
transfer.

From this, it is sensible to conclude that the variational analysis
presented here is more suited to studying dynamics at the onset of the
collision, before the soliton envelopes significantly overlap, but not in
the course of  the collision proper.

\subsubsection{\label{sec:level3}The interaction potential}

The set of coupled differential equations can be reformulated into a
mechanical system, in order to derive an effective potential describing the
interaction between the solitons. We begin by first restricting our analysis
at the onset of the collision, before the solitons began to significantly
overlap, as said above. In this regime, the second soliton remains
approximately stationary and we can fix that $\xi _{2}(t)=0$ and $v_{2}(t)=0$%
. The set of coupled differential equations then greatly simplify, and we
may readily integrate Eq. (\ref{difxi1}) to obtain
\begin{equation}
v_{1}\dot{\xi}_{1}\gamma =-4\left. \langle \mathcal{L}_{I}\rangle
\right\vert _{\dot{q}_{n}}+C,  \label{potint}
\end{equation}%
where $C$ is an arbitrary integration constant. Substituting Eq. (\ref{v1})
into Eq. (\ref{potint}), and then substituting again to remove a factor of $%
\dot{\xi}_{1}$, leads to the mechanical energy equation,
\begin{equation}
\frac{1}{2}\dot{\xi}_{1}^{2}+\frac{a_{1}}{6bf^{2}\gamma }\dot{\xi}_{1}+V_{%
\mathrm{int}}=\frac{C}{2f\gamma ^{2}},  \label{newton}
\end{equation}%
where we identify the soliton kinetic energy $\dot{\xi}_{1}^{2}/2$, total
energy $C/2f\gamma ^{2}$, and the effective interaction potential,
\begin{equation}
V_{\mathrm{int}}=-\frac{1}{2f^{2}\gamma ^{2}}\left[ d\left( d+\frac{a_{1}}{%
6bf}-v_{1}\right) -4f\left. \langle \mathcal{L}_{I}\rangle \right\vert _{%
\dot{q}_{n}}\right] .  \label{interaction potential}
\end{equation}%
The structure of Eq. (\ref{newton}), treats the motion of the first soliton
as a classical particle moving through the potential landscape of the second.

\begin{figure}[t]
\includegraphics[width=9.25cm]{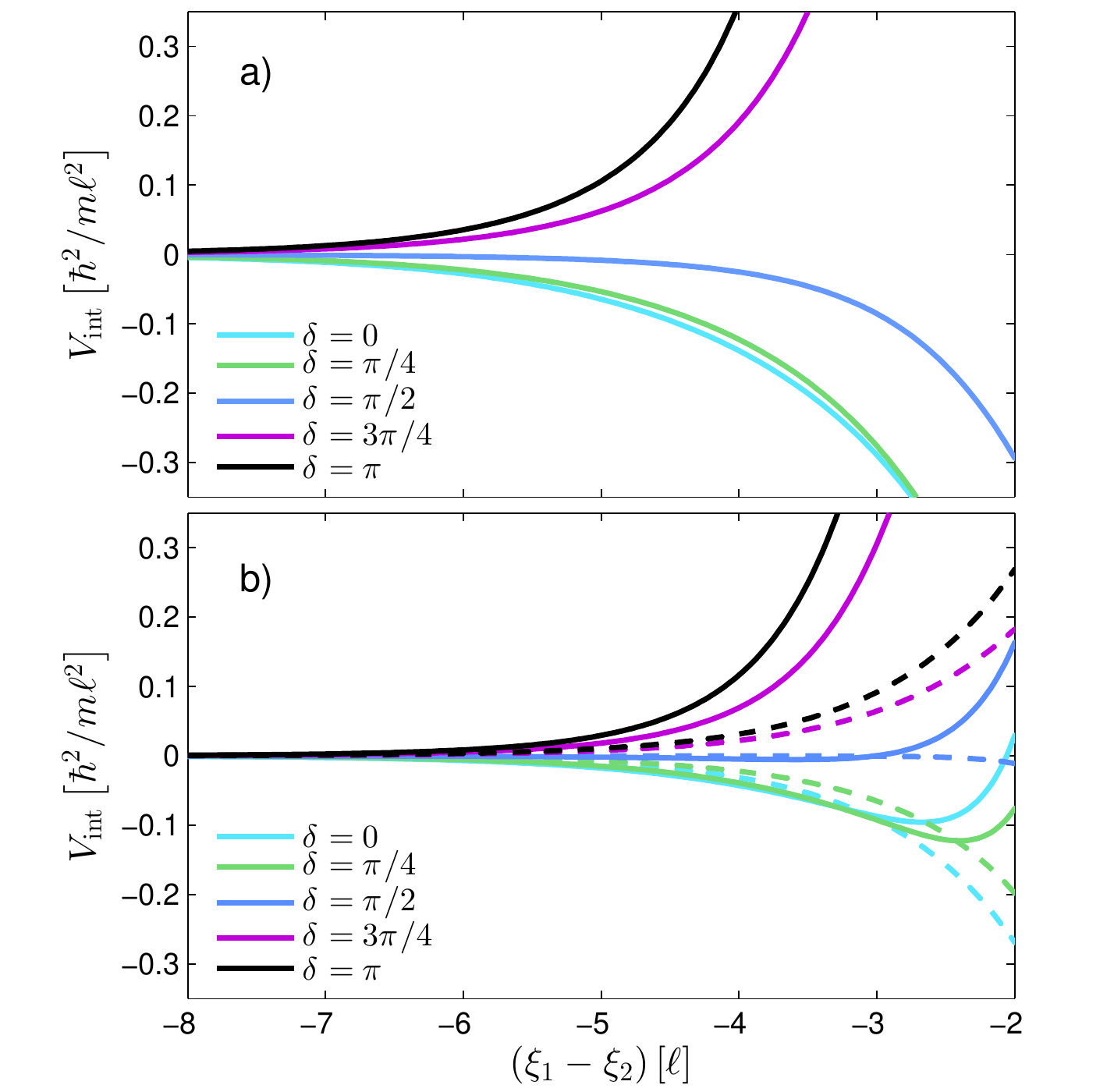}
\caption{(colour online). (a) Potential-energy curves, predicted by Eq. (%
\protect\ref{interaction potential}) as a function of the soliton
separation, for various phase differences. (b) Comparison between asymptotic
curves obtained by Eq. (\protect\ref{Lagasym}) (solid) and Eq. (\protect\ref%
{asymPOT}) (dashed). In both cases, the parameters are taken as $g_{\mathrm{%
1D}}=-4$, $v_{1}=0.1$, and $a_{1}=3$.}
\label{potpic}
\end{figure}

We plot the interaction curves in Fig. \ref{potpic}, in comparison to an
asymptotic calculation derived in the next section.
The interaction curves are here plotted only for negative values of the
separation, as we have considered the situation in which the moving soliton
approaches the stationary soliton up to the start of the collision.

For a typical set of parameters which we have used in our simulations, the
curves in Fig. \ref{potpic} show that both repulsive and attractive
interactions are supported, up to a choice of the phase difference. The
presence of an attractive potential---in particular, for $\delta =0$ and $%
\delta =\pi /4$,---therefore supports the existence of bound states, and the
interpretation of the `resonance' regime in Fig. \ref{solcol3} as an example
of the attractive dynamics. The effect of the divergence of the amplitude
for $\delta >\pi /2$ is illustrated by the upper two curves of Fig. \ref%
{potpic}, as the magnitude of the potential increases rapidly.

In fact, the nature of these curves does not differ drastically from the
results expected in the NLS equation, as contributions from terms stemming
from the current feature as a short-range attractive potential, hence they
do not have significant influence far from the centre of the interaction.
This can be compared to the plot with $a_{1}=0$, from which it is well known
that $\delta =0$, $\delta =\pi $, and $\delta =\pi /2$ correspond,
respectively, to the attractive, repulsive, and weak intermediate
interactions of solitons in the standard NLS equation.

\subsection{\label{Asymptotic}Asymptotic calculations}

In the previous section, the interaction of two chiral solitons was explored
using a full variational approach. It is further useful to consider an
asymptotic approach for well-separated solitons. Such an analysis can be
accomplished using the methodology presented in Refs.~\cite{boris1,boris2},
where tails of the solitons are used to produce the interaction potential (see also \cite{other_method}).
We begin by writing the one-dimensional Hamiltonian density for the system,
\begin{equation}
\mathcal{H}=-\frac{1}{2}|\left( \hat{p}-a_{1}|\psi |^{2}\right) \psi |^{2}-%
\frac{g_{\mathrm{1D}}}{2}|\psi |^{4},  \label{Hamildens}
\end{equation}%
which, when minimised, reduces to Eq. (\ref{mincoupleEOM}).
To derive an effective interaction potential, we again restrict our analysis to the regime of weak-chiral interactions in which the second soliton can be taken as stationary with respect to the first. In addition, since $|a_{1}(v_{1}+v_{2})|\ll |g|$, we note that the dominant contributions to the interaction potential in Eq.~(\ref{interaction potential}) arise from terms not containing a factor of $v_1$. Therefore we can further impose that the first soliton is also stationary such that $v_1=v_2=0$. Then provided the solitons' centres of masses are well separated by a distance $L=\xi_1-\xi_2$, we may approximate the two-soliton state in the vicinity of the first soliton as
\begin{equation}
\psi =\psi _{1}(x)+\psi _{2}(x+L),  \label{tail}
\end{equation}%
in which $\psi _{1}$ represents the envelope of the first soliton, and $\psi
_{2}$ is the exponential tail of the second soliton. The reciprocal
approximation is valid in the region around the second soliton.

Next, we substitute Eq.~(\ref{tail}) and its counterpart pertaining to the
second soliton into Eq.~(\ref{Hamildens}), and retain terms which are
linear with respect to the small tails, either $\psi _{1}$ or $\psi _{2}$.
The resulting expression can be recast in a compact form as
\begin{equation}
\mathcal{H}_{12}=\left. \left[ \psi _{2}\frac{\partial \mathcal{H}}{\partial
\psi }+\partial _{x}\psi _{2}\frac{\partial \mathcal{H}}{\partial \psi _{x}}+%
\mathrm{c.c.}\right] \right\vert _{\psi =\psi _{1}},  \label{Hamvary}
\end{equation}%
where each derivative is evaluated at point $\psi =\psi _{1}$.

To proceed, we note that the Hamiltonian can be defined as $H=\int_{-\infty
}^{+\infty }dx\mathcal{H}$, with the variational (alias Fr\'{e}chet)
derivative,
\begin{equation}
\frac{\delta H}{\delta \psi }=\frac{\partial \mathcal{H}}{\partial \psi }-%
\frac{\partial }{\partial x}\frac{\partial \mathcal{H}}{\partial \psi _{x}}.
\label{Hamfunc}
\end{equation}%
As we consider only the case of exact solutions for which $\delta H/\delta
\psi =0$, we may recast Eq. (\ref{Hamvary}), using Eq. (\ref{Hamfunc}) and
integration by parts, to obtain the expression
\begin{equation}
H_{12}=\left[ \psi _{2}\left. \frac{\partial \mathcal{H}}{\partial \psi _{x}}%
+c.c\right] _{z_{0}}^{\infty }\right\vert _{\psi =\psi
_{1}}+\{1\rightleftharpoons 2\},  \label{Hamnew}
\end{equation}%
which, after using Eq.~\eqref{Hamvary}, simplifies to
\begin{equation}
H_{12}=\left[ \left( \frac{1}{2}\partial _{x}\psi _{1}^{\ast }{+}\frac{ia_{1}%
}{2}\psi _{1}^{\ast }|\psi _{1}|^{2}\right) \psi _{2}+\mathrm{c.c.}\right]
_{z_{0}}^{\infty }{+}\{1\rightleftharpoons 2\}.  \label{12}
\end{equation}%
Rather than taking the integration limits in the domain $(-\infty ,\infty )$%
, we here divide the integration domain at an arbitrary point $z_{0}$
located between the solitons, and introduce the symmetric contribution $%
\{1\rightleftharpoons 2\}$ to account for the contribution from the second
soliton.
To proceed, we require expressions for single-soliton states of Eq. (\ref%
{tail}), which are given by
\begin{equation}
\psi _{n}=\frac{1}{2\sqrt{b}}e^{i\phi _{n}}\times
\begin{cases}
\sech(x/b), & n=1, \\
\sech((x+L)/b), & n=2,%
\end{cases}
\label{two-sol}
\end{equation}%
where the soliton's width $b=-4/\tilde{g}$ is chosen such that $%
\sum_{n}\int_{-\infty }^{+\infty }dx|\psi _{n}|^{2}=1$, with each soliton
containing half the number of atoms. In keeping with the linearisation
procedure used in deriving Eq. (\ref{Hamnew}), we can instead simplify these
expressions in the vicinity of $z_{0}$, with asymptotic forms
\begin{equation}
\psi _{n}\sim \frac{1}{\sqrt{b}}e^{i\phi _{n}}\times
\begin{cases}
e^{-x/b}, & n=1, \\
e^{\left( x+L\right) /b}, & n=2.%
\end{cases}
\label{asymforms}
\end{equation}%
To calculate the variational derivative in Eq. (\ref{Hamfunc}), we use the
full expressions in Eq. (\ref{two-sol}) to evaluate the upper limit at $%
x=\infty $, together with the asymptotic forms in Eq. (\ref{asymforms}) to
evaluate the lower limit at $z_{0}$. To obtain a contribution from the
current to the effective potential, which is independent of the choice of
arbitrary point $z_{0}$, we must go to the next order in the expression for $%
\psi _{2}$ in the second term in expression (\ref{12}), taking $\psi
_{2}\sim -\sqrt{2}e^{3\left( x+L\right) /b}e^{i\phi _{2}}/\sqrt{b}$.
Substituting these expressions, we obtain an effective interaction potential
\begin{equation}
V_{\mathrm{int}}=-\frac{2}{b^{2}}\left[ e^{\epsilon }\cos (\delta
)+a_{1}e^{3\epsilon }\sin (\delta )\right] ,  \label{asymPOT}
\end{equation}%
which is convenient to extract information from. The first term in Eq. (\ref%
{asymPOT}) comes from the NLS equation, being attractive/repulsive for the
correct choice of $\delta $. The second term, which appears due
to the current, may also be attractive or repulsive, but it is out of phase
with the first term. However this term yields a shorter interaction range,
compared to the NLS term, hence it does not contribute significantly to the
interaction potential far from the centre of the interaction.

It is relevant to compare the interaction potentials produced by both the
linear-superposition and asymptotic models. Because the asymptotic
calculation neglects some terms in the underlying Hamiltonian density,
compared to the linear-superposition ansatz (\ref{Solansatz}), it is
necessary to approximate the result produced by the linear superposition,
using the asymptotic forms, so that they can be fairly compared. This is
detailed in Appendix \ref{asymptoticpot}. Figure \ref{potpic} shows a
comparison between these interaction potentials, from which we can see that
both curves share the same qualitative features, with both repulsive and
attractive interactions supported in a similar manner to the standard NLS
equation.

\section{\label{Conclusion}Conclusion}

We have demonstrated how the current nonlinearity introduces non-integrable
effects in the collision dynamics of bright matter-wave solitons. Using the
variational approximation, we have derived an effective particle model for
the soliton dynamics, which helps to explain both inelastic scattering and
the attractive/repulsive nature of the interactions. We have also derived
effective potentials for the  interaction between the solitons. We showed
that the particle model is valid as long as the current nonlinearity is
weak, similar to the situation in integrable models, where inherent
symmetries of the system can be exploited \cite{solint3}. This fact implies
that essential results may be produced by collisions between slowly moving
solitons. For stronger interactions, the particle model breaks down due to
the non-integrability. We observe, in particular, how the strong
current-induced nonlinearity can destabilize bound states of solitons and
also induce soliton fission, breaking two colliding solitons into several
ones after the collision.

These concepts constitute a rich spectrum of dynamics which are interesting
from the point of view of the fundamental nonlinear dynamics. In addition,
the chiral properties of the quantum gas studied here may provide novel
applications to atomtronics \cite{atomtronics} and quantum transport. In
such scenarios, careful consideration of the collision dynamics is needed.

\section{\label{sec:level1}Acknowledgements}

R.J.D. acknowledges support from EPSRC CM-CDT Grant No. EP/L015110/1, M.J.E.
acknowledges support from EPSRC Grant No. EP/M005127/1 and as an Overseas
researcher under Postdoctoral Fellowship of Japan Society for the Promotion
of Science, and P.\"{O}. acknowledges support from EPSRC grant No.
EP/M024636/1. B.A.M. appreciates hospitality of the Institute of Photonics
and Quantum Sciences at the Heriot-Watt University. The work of that author
is supported, in part, by the joint program in physics between NSF and
Binational (US-Israel) Science Foundation through project No. 2015616, and
by the Israel Science Foundation through Grant No. 1286/17.

\appendix

\section{Calculation of interaction integrals}

\label{integrals}

In this appendix we show how to evaluate the interaction integrals which
appear in the variational analysis, using the method of residues. These
calculations can be found in the standard literature \cite{solint1, solint3}%
, but we recapitulate details here, to provide a basis for evaluating more
complicated integrals.

\subsection{\label{app:integrals}Integral example I}

\begin{figure}[t]
\includegraphics[width=8.75cm]{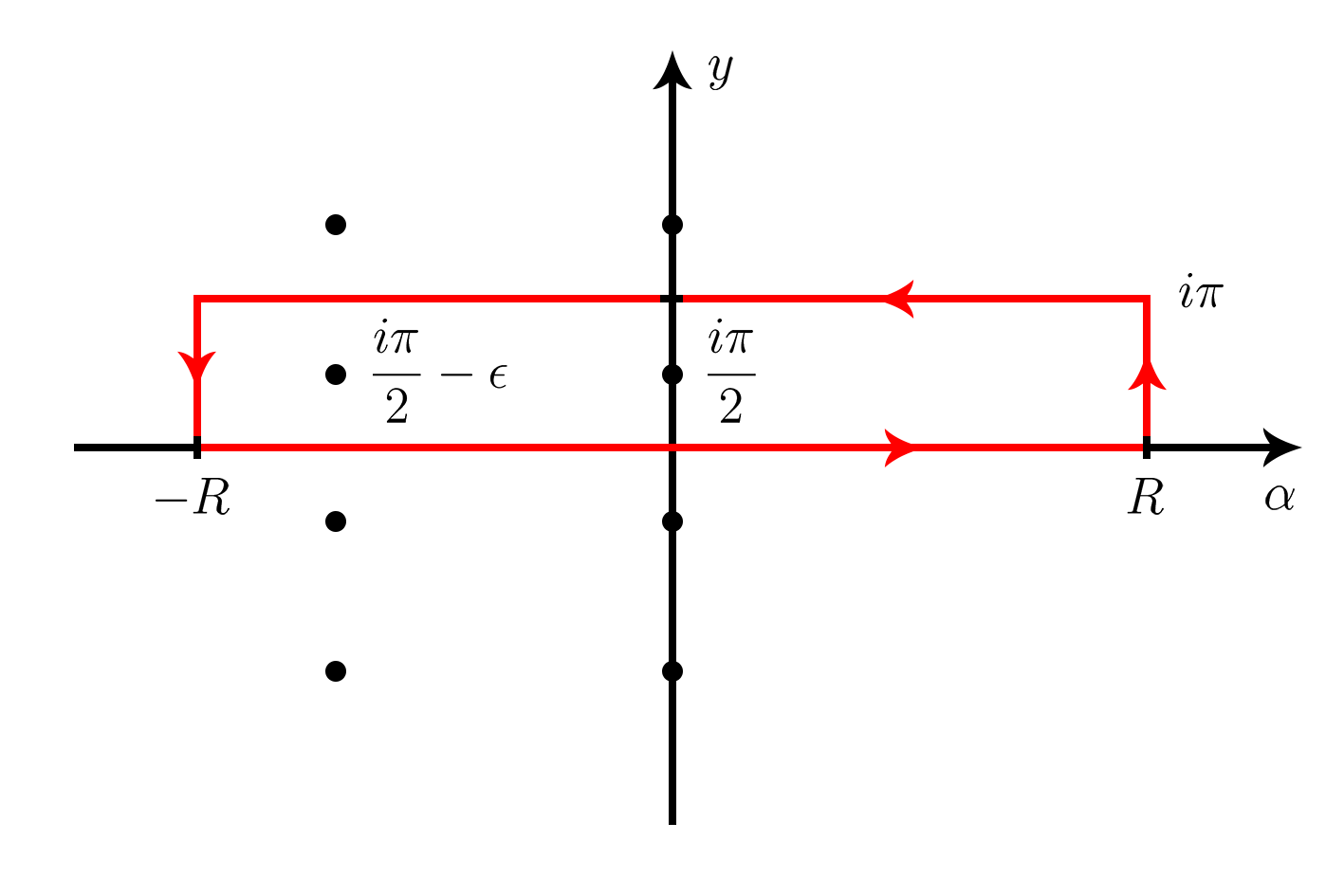}
\caption{(colour online). Map of the complex plane $z=\protect\alpha +iy$,
highlighting the location of the poles (black dots) and contour path $%
\mathcal{C}$ (red line) for all interaction integrals. The integration path
is counter-clockwise.}
\label{contour}
\end{figure}

The simplest interaction integral to evaluate is
\begin{equation}
I_{1}=b\int_{-\infty }^{+\infty }\frac{d\alpha }{\cosh (\alpha )\cosh
(\alpha +\epsilon )}.  \label{Int1}
\end{equation}%
To proceed, we consider the following contour integral
\begin{equation}
\oint_{\mathcal{C}}f(z)dz=\oint_{\mathcal{C}}\frac{z}{\cosh (z)\cosh
(z+\epsilon )}dz,
\end{equation}%
in which the contour path $\mathcal{C}$ forms a rectangular region in the
complex plane, $z=\alpha +iy$, with dimensions $-R<\alpha <R$, $0<y<\pi $.
The complex function $f(z)$ is analytic in the region except for a pair of
(simple) poles at $z_{1}=i\pi /2$ and $z_{2}=i\pi /2-\epsilon $. These
properties are illustrated in Fig. (\ref{contour}).

In the limit of $R\rightarrow \infty $, the contour integrals along the
vertical paths vanish, as $f(z)$ exponentially converges to zero at $z=\pm
\infty $. The horizontal paths also cancel, except for a contribution from
the top path which is proportional to the desired integral. Therefore, from
the residue theorem we can write
\begin{equation}
b\oint_{\mathcal{C}}f(z)dz=-i\pi I_{1}=2\pi ib\sum_{k=1,2}\text{Res}\left(
f(z),z_{k}\right) .
\end{equation}%
The task of evaluating $I_{1}$ reduces to simply, albeit tediously,
computing the residues of $f(z)$, which are given by
\begin{equation}
\begin{split}
\text{Res}\left( f(z),z_{1}\right) & =\lim_{z\rightarrow z_{1}}\left(
z-z_{1}\right) f\left( z\right)  \\
& =\frac{\pi }{2i\sinh (\epsilon )},
\end{split}%
\end{equation}
\begin{equation}
\begin{split}
\text{Res}\left( f(z),z_{2}\right) & =\lim_{z\rightarrow z_{2}}\left(
z-z_{2}\right) f\left( z\right)  \\
& =-\frac{\pi }{2i\sinh (\epsilon )}-\frac{\epsilon }{\sinh (\epsilon )}.
\end{split}%
\end{equation}%
Collating the results, we find
\begin{equation}
I_{1}=\frac{2b\epsilon }{\sinh (\epsilon )}.
\end{equation}

\subsection{\label{app:subsec}Integral example II}

A more involved example is
\begin{equation}
I_{2}=b\int_{-\infty }^{\infty }\frac{\alpha \tanh (\alpha +\epsilon )}{%
\cosh (\alpha )\cosh (\alpha +\epsilon )}d\alpha .  \label{Int2}
\end{equation}%
In this case, we consider the following contour integral:
\begin{equation}
\oint_{\mathcal{C}}g(z)dz=\oint_{\mathcal{C}}\frac{z^{2}\sinh (z+\epsilon )}{%
\cosh (z)\cosh ^{2}(z+\epsilon )}dz,
\end{equation}%
in which the contour path $\mathcal{C}$ is identical to the one used in the
previous example. In contrast, the integrand now contains a simple pole and
a second-order one. Again, using the residue theorem, we can write
\begin{equation}
b\oint_{\mathcal{C}}g(z)dz=-2\pi iI_{2}+\pi ^{2}I_{3}=2\pi ib\sum_{k=1,2}%
\text{Res}\left( g(z),z_{k}\right) ,
\end{equation}%
where the task of evaluating $I_{2}$ now depends on computing the residues
of $g(z)$ and knowledge of the integral
\begin{equation}
\begin{split}
I_{3}& =b\int_{-\infty }^{\infty }\frac{\tanh (\alpha +\epsilon )d\alpha }{%
\cosh (\alpha )\cosh (\alpha +\epsilon )} \\
& =-\frac{2b}{\sinh (\epsilon )}+\frac{2b\epsilon \cosh (\epsilon )}{\sinh
^{2}(\epsilon )}.
\end{split}
\label{Int17}
\end{equation}%
The residues in this instance are given by
\begin{equation}
\begin{split}
\text{Res}\left( g(z),z_{1}\right) & =\lim_{z\rightarrow z_{1}}\left(
z-z_{1}\right) g\left( z\right)  \\
& =\frac{\pi ^{2}\cosh (\epsilon )}{4\sinh ^{2}(\epsilon )},
\end{split}%
\end{equation}
\begin{equation}
\begin{split}
\text{Res}\left( g(z),z_{2}\right) & =\lim_{z\rightarrow z_{2}}\frac{d}{dz}%
\left[ \left( z-z_{2}\right) ^{2}g\left( z\right) \right]  \\
& =(i\pi -2\epsilon )\frac{1}{\sinh (\epsilon )}+\left( \epsilon -\frac{i\pi
}{2}\right) ^{2}\frac{\cosh (\epsilon )}{\sinh ^{2}(\epsilon )}.
\end{split}%
\end{equation}%
Collating the results, we find
\begin{equation}
I_{2}=-\frac{b\epsilon ^{2}\cosh (\epsilon )}{\sinh ^{2}(\epsilon )}+\frac{%
2b\epsilon }{\sinh (\epsilon )}.
\end{equation}%
For the remaining interaction integrals, one can proceed using the same
methodology, provided that the correct contour shift is applied to the
integral, namely,

\begin{equation}
\int^{\infty}_{-\infty}f\left(\alpha,\epsilon\right)\alpha^{n-1}d\alpha%
\rightarrow\int_{\mathcal{C}}f\left(z,\epsilon\right)z^{n}dz,
\end{equation}
or
\begin{equation}
\int^{\infty}_{-\infty}f\left(\alpha,\epsilon\right)\left(\alpha+\epsilon%
\right)^{n-1}d\alpha\rightarrow\int_{\mathcal{C}}f\left(z,\epsilon\right)%
\left(z+\epsilon\right)^{n}dz,
\end{equation}
where $n\in\mathbb{Z}^+$ and $f(\alpha,\epsilon)$ is a product of hyperbolic
functions.\newline

\section{The asymptotic approximation for the interaction potential produced
by the linear-superposition ansatz}

\label{asymptoticpot}

To directly compare the interaction potentials derived from the variational
analysis, it is necessary to reduce the potential obtained from the
linear-superpositon ansatz (\ref{Solansatz}) to a form which captures the
same approximations which were used in the asymptotic version, based on
ansatz (\ref{tail}). This can be done by introducing asymptotic forms for
each contribution to the interaction potential and selectively dropping
terms which are either short-range ones, or contain an explicit dependence
on the soliton velocities.

We start, by explicitly restating the linear interaction potential
\begin{widetext}
\begin{equation}\label{lalala}
\begin{split}
V_{\rm int}&=-\frac{1}{2(bf\gamma)^2}\left(d^2+\frac{a_1}{6bf}d\right)+\frac{g}{bf\gamma^2}\left(\left(1+\cos(2\delta)/2\right)\left[\frac{\cosh(\epsilon)}{\sinh^3(\epsilon)}-\frac{1}{\sinh(\epsilon)}\right]+\cos(\delta)\left[\frac{\cosh(\epsilon)}{\sinh^2(\epsilon)}-\frac{\epsilon}{\sinh^3(\epsilon)}\right]\right)\\
&+\frac{1}{b^2}\cos(\delta)\left[\frac{\cosh(\epsilon)}{\sinh^2(\epsilon)}+\frac{\epsilon}{\sinh(\epsilon)}-\frac{\epsilon\cosh^2(\epsilon)}{\sinh^3(\epsilon)}\right],
\end{split}
\end{equation}
\end{widetext}which has been simplified by taking $v_{1}=0$, for the case of
two stationary solitons. To reduce Eq. (\ref{lalala}) to an asymptotic form,
we recall that, while deriving Eq. (\ref{interaction potential}), we
considered the dynamics of the soliton travelling in the positive $x$%
-direction under the action of the effective potential induced by the second
soliton. Therefore, we are required to approximate the interaction potential
for negative values of $\epsilon $, with respect to the second soliton which
is centered at $x=0$. Then, in the same manner as before, we introduce the
following asymptotic forms: $\cosh (\epsilon )\sim e^{-\epsilon }/2$, $\csch%
(\epsilon )\sim -2\epsilon ^{\epsilon }$, $f(\epsilon ,\delta )\sim
1-2e^{\epsilon }\epsilon \cos (\delta )$, and
\begin{equation}
\gamma (\epsilon ,\delta )\sim 1-\frac{\left( 1-f\right) ^{2}\left(
1+\epsilon \right) -f\left( 1-f\right) \left( 2+\epsilon \right) }{2f^{2}}.
\end{equation}%
Substituting these expressions into Eq. (\ref{lalala}), one arrives at the
asymptotic potential
\begin{widetext}
\begin{equation}\label{Lagasym}
\begin{split}
V_{\rm int}&\sim-\frac{2}{\left(bf\gamma\right)^2}\left(-\sin(\delta)\left(\epsilon+1\right)e^\epsilon-\frac{2a_1}{f}\left(1+\cos(2\delta)/2\right)\left(\epsilon+1\right)e^{2\epsilon}+\frac{a_1}{f\hbar}\cos(\delta)e^\epsilon\right)^2\\
&+\frac{2}{f\gamma^2}\left(-\frac{4g}{b}\left(1+\cos(2\delta)/2\right)e^{2\epsilon}+\frac{2}{b^2}e^\epsilon\cos(\delta)+\frac{g}{b}\cos(\delta)e^\epsilon\right).
\end{split}
\end{equation}
\end{widetext}In writing Eq. (\ref{Lagasym}), we have neglected term $%
\propto e^{3\epsilon }$ and the term $\propto a_{1}d\left( \epsilon ,\delta
\right) /6$ in Eq. (\ref{lalala}), as they, being small in the asymptotic
limit, do not essentially affect the result.

\bibliographystyle{iopart-num}
\providecommand{\noopsort}[1]{}\providecommand{\singleletter}[1]{#1}%
\providecommand{\newblock}{}

\end{document}